\documentclass[preprint,11pt]{aastex}

\newcommand{\wfu}{\hbox{$U_{300}$}}
\newcommand{\wfb}{\hbox{$B_{450}$}}
\newcommand{\wfv}{\hbox{$V_{606}$}}
\newcommand{\wfi}{\hbox{$I_{814}$}}
\newcommand{\nicj}{\hbox{$J_{110}$}}
\newcommand{\nich}{\hbox{$H_{160}$}}

\newcommand{\mathM}{\hbox{$\mathcal{M}$}}

\newcommand{\etal}{et al.}

\slugcomment{To appear in the Astrophysical Journal}

\shorttitle{Evolution of the Global Stellar Mass Density}
\shortauthors{Dickinson, Papovich, Ferguson \& Budav\'{a}ri}

\received{2002 May 30}
\begin{document}

\title{THE EVOLUTION OF THE GLOBAL STELLAR MASS DENSITY\\
 AT \mbox{\boldmath $0 < z < 3$}\altaffilmark{1}}

\author{{\sc Mark Dickinson}\altaffilmark{2,5}}
\email{med@stsci.edu}

\author{{\sc Casey Papovich}\altaffilmark{3,4}}
\email{papovich@as.arizona.edu}

\author{{\sc Henry C. Ferguson}\altaffilmark{2}}
\email{ferguson@stsci.edu}

\and
\author{{\sc Tam\'{a}s Budav\'{a}ri}\altaffilmark{4}}
\email{budavari@pha.jhu.edu}

\altaffiltext{1}{Based on observations taken with the NASA/ESA Hubble
Space Telescope, which is operated by the Association of Universities
for Research in Astronomy, Inc.\ (AURA) under NASA contract
NAS5--26555}  

\altaffiltext{2}{Space Telescope Science Institute, 3700 San Martin Dr., 
Baltimore, MD 21218}

\altaffiltext{3}{Steward Observatory, University of Arizona, Tucson, AZ 85721}

\altaffiltext{4}{Department of Physics and Astronomy, The Johns Hopkins 
University, Baltimore, MD 21218}

\altaffiltext{5}{Visiting Astronomer, Kitt Peak National Observatory, 
National Optical Astronomy Observatories, which is operated by the 
Association of Universities for Research in Astronomy, Inc.  (AURA) 
under cooperative agreement with the National Science Foundation.}

\begin{abstract}

The build--up of stellar mass in galaxies is the consequence of their past 
star formation and merging histories.  Here we report measurements of 
rest--frame optical light and calculations of stellar mass at high redshift 
based on an infrared--selected sample of galaxies from the Hubble Deep Field 
North.   The bright envelope of rest--frame $B$--band galaxy luminosities 
is similar from $0 < z < 3$, and the co--moving luminosity density is constant 
to within a factor of 3 over that redshift range.  However, galaxies at higher 
redshifts are bluer, and stellar population modeling indicates that they had 
significantly lower mass--to--light ratios than those of present--day $L^\ast$ 
galaxies.  This leads to a global stellar mass density, $\Omega_\ast(z)$, which 
rises with time from $z = 3$ to the present.  This measurement essentially traces 
the integral of the cosmic star formation history that has been the subject of 
previous investigations.  50--75\% of the present--day stellar mass 
density had formed by $z \sim 1$, but at $z \approx 2.7$ we find only 3--14\% of 
today's stars were present.  This increase in $\Omega_\ast$ with time is broadly 
consistent with observations of the evolving global star formation rate once dust 
extinction is taken into account, but is steeper at $1 < z < 3$ than predicted by 
some recent semi--analytic models of galaxy formation.  The observations appear 
to be inconsistent with scenarios in which the bulk of stars in present--day 
galactic spheroids formed at $z \gg 2$.   

\end{abstract}
 

\section{Introduction
\label{section:intro}}

One goal of observational cosmology is to understand the history
of mass assembly in galaxies.  In current models of structure formation, 
dark matter halos build up in a hierarchical process controlled by the 
nature of the dark matter, the power spectrum of density fluctuations, 
and the parameters of the cosmological model.  The assembly of the 
stellar content of galaxies is governed by more complex physics, 
including gaseous dissipation, the mechanics of star formation itself, 
and the feedback of stellar energetic output on the baryonic material 
of the galaxies.

While it is clear that the mean stellar mass density of the universe,
$\Omega_\ast$, should start at zero and grow as the universe ages,
the exact form of this evolution is not trivial, and could reveal 
much about the interaction of large--scale cosmological physics 
and small--scale star formation physics.  For example, there may 
have been multiple epochs of star formation in the universe, associated 
with changes in the ionizing radiation background.  The initial mass
function (IMF) may have been different for early generations of stars
or in different environments (e.g., in starbursts versus quiescent disks).
This would change the relation between luminosity, metal production
and total star formation rates.  Early star formation with a top--heavy IMF
might leave little in the way of remaining stellar mass, while contributing
significantly to the ionization of the intergalactic medium and dispersing
metals into it.   The total, integrated mass in stars 
is tightly coupled to the history of star formation as traced by the 
infrared background, and to the cold gas content of the universe, 
$\Omega_{\mathrm{HI}}$.  Reducing the uncertainties on all of these 
measurements will provide strong constraints on models for 
galaxy formation, while discrepancies could reveal new physics.

Deep imaging and spectroscopic surveys now routinely find and study 
galaxies throughout most of cosmic history, back to redshifts $z = 3$ 
and earlier.  A variety of studies have considered the evolution 
of the global star formation rate (i.e., averaged over all 
galaxies), using different observational tracers of star formation in 
distant galaxies \citep{lil96, mad96, mad98, ste99, bar00, tho01}, or
constraints from neutral gas content and/or the extragalactic background
light \citep{pei95, pei99, cal99}.  It is generally accepted that star 
formation in galaxies was more active at high redshift than today.  
Several investigations have found an approximately constant rate of global 
star formation from $z \approx 4$ to 1, with a steeper decline since
that epoch to the present--day, although other studies have 
questioned these results \citep{cow99, wil02, lan02}.  The true rates
of star formation (either globally or for individual galaxies)
are subject to a variety of uncertainties and biases, particularly 
regarding dust extinction and its effect on galaxy selection and on
observational indicators of star formation.

A complementary approach is to try to measure the stellar {\it masses}
($\mathM$) of galaxies, rather than the {\it rates} at which they are 
forming stars.  This is, at least indirectly, a common motivation for 
near--infrared galaxy surveys, which observe starlight that provides 
a reasonable tracer of the total stellar mass.  The stellar mass 
of a galaxy is dominated by older, lower luminosity, redder stars.  
At longer wavelengths, these stars provide a greater fractional 
contribution to the integrated light from the galaxy, relative 
to that from short--lived, massive stars.  For a stellar population
formed over an extended period of time, the luminosity--weighted 
mean age increases at redder wavelengths, and the range of possible 
mass--to--light ratios ($\mathM/L$) is reduced.  Moreover, the impact 
of dust extinction is greatly reduced at longer wavelengths.  

Measurements of the $K$--band luminosity function of nearby galaxies 
have been aimed, in part, at assessing the distribution
of stellar mass among galaxies at the present day.
For example, in a recent study, \citet{col01} combined data from 
the 2dF Galaxy Redshift Survey (2dFGRS) and the Two Micron All--Sky Survey 
(2MASS) to produce a large and well--controlled sample for determining 
the $K$--band luminosity function.  They employed stellar population synthesis 
models to relate light to mass, using the colors of galaxies to constrain
$\mathM/L$, and hence to derive the stellar mass function of galaxies in 
the nearby universe.   \citet{kau02} have recently used $z$--band 
(0.9$\mu$m) photometry and spectroscopic stellar population indices 
in order to evaluate the stellar mass content of a large sample of 
galaxies from the Sloan Digital Sky Survey (SDSS).

Similar methods have been applied to galaxies at cosmological
distances.  Deep near--infrared surveys are generally intended to 
measure the rest--frame optical light from high redshift galaxies, 
in order to minimize the effects of active star formation and dust on 
galaxy selection and to more nearly trace their stellar masses.
\citet{bri00} have used $K$--band photometry of galaxies from
redshift surveys of {\it Hubble Space Telescope (HST)} imaging fields 
to derive stellar masses for galaxies out to $z \approx 1$, studying the 
evolution of the global stellar mass density as a function of galaxy 
morphology.  \citet{coh02} has also combined extensive redshift survey
data with infrared photometry and evolutionary models to study the
evolving stellar mass density out to $z \approx 1$.  \citet{dro01} 
have used an infrared imaging and photometric redshift survey, 
which covers a larger solid angle to shallower depths, to study the 
distribution function of stellar mass at the bright end of the galaxy 
population over a similar redshift range.  
At higher redshift, \citet{saw98}, \citet{pap01} and \citet{sha01} have 
used deep infrared data to study the stellar mass content of Lyman Break 
Galaxies (LBGs;  cf., Steidel \etal\ 1996), actively star--forming galaxies 
at $z \sim 3$ that are selected on the basis of their UV luminosity and 
the broad band color signature of the Ly~$\alpha$ and Lyman limit breaks.

Here, we will follow the basic approach of \citet{col01}, \citet{bri00}
and \citet[henceforth PDF]{pap01} to determine stellar masses for 
galaxies in the Hubble Deep Field North (HDF--N) out to $z = 3$, and 
to study the redshift evolution of the global stellar mass density.
Throughout this paper, we assume a cosmology with 
$\Omega_{\rm tot}, \Omega_M, \Omega_\Lambda = 1.0, 0.3, 0.7$,
and $h \equiv H_0 / (100~\mathrm{km~s^{-1}~Mpc^{-3}}) = 0.7$.
We will sometimes write 
$h_{70} \equiv H_0 / (70~\mathrm{km~s^{-1}~Mpc^{-3}})$.
We use AB magnitudes 
($AB \equiv 31.4 - 2.5\log\langle f_\nu / \mathrm{nJy} \rangle$),
and denote the bandpasses used for {\it HST}
WFPC2 and NICMOS imaging as \wfu, \wfb, \wfv, \wfi, \nicj, and \nich.

\section{The HDF--N data
\label{section:data}}

The central region of the HDF--N which we analyze here covers 
a solid angle of approximately 5.0~arcmin$^2$.  We use deep imaging and 
photometric data in seven bandpasses spanning the wavelength range 0.3--2.2$\mu$m,
taken with {\it HST}/WFPC2 \citep{wil96}, {\it HST}/NICMOS \citep{dic99, dic00a, 
dic00b, pap01, dic03}, and at $K$--band with IRIM on the KPNO 4m 
\citep{dic98}.  The galaxy catalog was selected from a weighted
sum of the near--infrared NICMOS images using SExtractor
\citep{ber96}, and is highly complete to $\nich \approx 26.5$.  
The point spread functions of the other {\it HST} images were matched 
to that of the NICMOS data, and photometry was measured through 
matched apertures.  The ground--based $K$--band fluxes were 
measured using the TFIT method described in PDF.  ``Total'' 
magnitudes are approximated by measurements in elliptical apertures 
defined by moments of the light profile (SExtractor MAG\_AUTO).
In cases where neighboring objects penetrate the elliptical 
apertures, photometric contamination was handled by symmetric
replacement of pixels within the ``intruder'' by values drawn from 
the opposite side of the galaxy being measured.  The colors
used for the photometric redshift and spectral fitting analyses
described below were measured through isophotal apertures defined
in the summed NICMOS detection image and fixed for the other bands.

The most complete collection of HDF--N 
redshifts known to the authors was employed, taken primarily 
from \citet{coh00}, \citet{coh01}, and \citet{daw01}, 
and references therein, plus a few additional redshifts made 
available by K.\ Adelberger and C.\ Steidel (private communication).
Spectroscopic redshifts are used whenever possible, but these only
cover the $\sim 170$ brightest galaxies in the sample.  For the
remainder, we use photometric redshift estimates fit to our 
catalog by \citet{bud00}.  Where comparison is possible,
these photometric redshifts give excellent agreement with the 
spectroscopic values, with 91\% of objects (at all redshifts,
$0 < z < 6$) having 
$|\Delta_z| \equiv |z_{phot} - z_{spec}| / (1 + z_{spec}) < 0.1$,
and only 5\% ``catastrophic outliers'' with $|\Delta_z| > 0.2$.
\citet{coh00} have reported lower quality or confidence classes 
for some of these outliers.  Many of them were also noted by \citet{fer01}, 
and in some cases the spectroscopic redshifts were subsequently 
updated by \citet{coh01} or \citet{daw01}.  In a few cases where 
we strongly suspect that a discrepancy remains (\citet{wil96} catalog
objects 3-355, $z_{spec}=0.474$; 4-316, $z_{spec}=2.801$; 
4-445, $z_{spec}=2.500$), we have adopted our photometric redshift 
estimates ($z_{phot} = 1.20$, 1.77 and 1.44, respectively) over the 
reported spectroscopic values (see also PDF for a discussion of the 
two higher redshift objects).  

The RMS fractional redshift difference for 
objects with $|\Delta_z| < 0.1$ is $\sigma(\Delta_z) = 0.031$ at 
$z < 1.9$, and $\sigma(\Delta_z) = 0.038$ at $z > 1.9$.  We cannot 
be certain that this uncertainty or the error rate does not increase 
at fainter magnitudes, below the spectroscopic limit, but as we will 
see, most of the luminosity and mass density in the HDF is 
contained in the more luminous objects, so photometric redshift 
errors near the catalog limits should not play a major role.  
We will discuss the importance of photometric redshift uncertainties
in \S\ref{section:photzerrs}.

Ten spectroscopically identified galactic stars, plus five other, 
fainter point sources with colors (primarily $\nicj-\nich$ vs.\
$\wfv-\wfi$) similar to those of the known stars, have been eliminated 
from the catalog.  Because we are interested in the rest--frame optical 
light, we limit our sample to galaxies at $z \leq 3$, where the deep 
\nich\ images measure rest--frame wavelengths of 0.4$\mu$m or redder.
Altogether, our HDF--N sample consists of 737 galaxies with 
$\nich \leq 26.5$ and $z \leq 3$. 

\section{Stellar mass estimates
\label{section:masses}}

In order to estimate the stellar masses of HDF--N galaxies, we fit 
their broad--band photometry using model spectra generated using 
a recent (2000) version of the stellar population synthesis code 
of Bruzual \& Charlot \citep{bru93, bru00}.  The procedure used
here is identical to that described in PDF, to which the reader is
directed for a full account.  It is also conceptually similar to the
methods used by other studies at lower redshifts to which we will 
compare our results.  Very briefly, we generate
a large suite of model spectra, varying the model age, the time scale of 
star formation (from instantaneous bursts through constant rates), and 
the dust extinction (parameterized using the ``starburst'' attenuation 
law of \citet{cal00}).   Other parameters, which cannot be well constrained 
(if at all) from photometry alone, are held fixed.  We discuss some
of these parameters below.   We fit each model spectrum to the galaxy 
photometry, excluding bands which fall below Ly~$\alpha$.  We select the 
best model based on the $\chi^2$ statistic, and use Monte Carlo simulations 
to define the 68\% confidence interval in our model parameters by perturbing
the galaxy photometry randomly within its error distribution.

PDF examined the dependence of the derived stellar population parameters 
for $z > 2$ LBGs on the form of the extinction law.  While changing from 
the flat starburst law to a steep SMC extinction curve had a strong impact 
on some of the derived quantities, such as ages or star formation rates, 
the distribution of LBG stellar masses was virtually unchanged.  Therefore, 
we have not further explored this as a variable in the present investigation.  
We have also fit models with different (but fixed) metallicities ($Z$).  
Here, we will use results from models with $Z = 0.02$ (i.e., approximately 
solar) and 0.004 to estimate the range of systematic uncertainty due 
to metallicity effects.  

Any conversion of light to stellar mass requires certain assumptions, most 
notably about the initial mass function (IMF) of star formation.  In the 
present analysis, we adopt a \citet{sal55} IMF with lower and upper mass 
cutoffs of 0.1 and 100~$\mathM_\odot$, which allows us to compare our 
results to others from the literature for galaxies at lower redshifts.  
To first order, varying the IMF simply scales our derived $\mathM/L$ 
up or down.  Adopting an IMF with a larger low--mass cutoff or
flatter low--mass slope would result in smaller derived galaxy masses 
for the same amount of optical or near--infrared light.  This difference 
is a factor of $\sim 2$ for the IMFs of \citet{ken83} or \citet{kro01}.
However, the evolution of stellar mass in galaxies, which we discuss here, 
should be largely independent of the faint--end slope of the IMF, 
provided that the IMF itself does not evolve.   Changing the IMF 
at intermediate to high stellar masses, however, can introduce
systematic differences in the evolution of $\mathM/L(\lambda_0)$ with stellar 
population age and as a function of the rest--frame wavelength $\lambda_0$.
Note that both stellar masses and star formation rates derived from 
photometric observations depend strongly on assumptions about the IMF, 
but in somewhat different ways.  Most observational indicators of star 
formation are sensitive only to very high mass stars, requiring large 
and IMF--dependent extrapolations to the ``total'' star formation rate.
Rest--frame infrared and optical photometry generally measures light 
from stars spanning a broader range of masses.  We will touch on issues
related to the IMF again in \S\ref{section:discuss_evol}.

As discussed in PDF, the range in $\mathM / L$ for models which 
fit the HDF--N galaxy photometry depends on the assumptions about
their previous star formation histories, $\Psi(t)$.  As in that 
paper, we bound the range of possible star formation histories 
by using two sets of models.  In one, we parameterize 
$\Psi(t) \propto \exp(-t/\tau)$, with time scale $\tau$ and 
age $t$ as free parameters.  These models span the range from 
instantaneous bursts to constant star formation.  The model ages 
are restricted to be younger than the age of the Universe at the 
redshift of each galaxy being fit.  Since very young ages with active 
star formation and no dust are allowed, this set of models includes 
the minimum permissible $\mathM / L$ for each galaxy.  The second 
set of models sets an upper bound on the stellar mass by adopting a rather 
artificial, two--component star formation history.  The ``young'' 
component is modeled with an exponential $\Psi(t)$ as described 
previously.  The ``old'' component is formed in an instantaneous 
burst at $z = \infty$, providing a maximally old population.
Given its age ($> 2$~Gyr for $z < 3$), this old component is 
assumed to be unreddened.  In principle, arbitrarily large masses
could be invoked by freely adding extinction to the older 
component, but we do not consider such models here, and will
refer to our 2--component models has having the ``maximum $\mathM/L$'' 
consistent with the photometry.  More complex star formation 
histories are bracketed by results from these two sets of models.

For LBGs at $z \sim 3$, PDF found that vigorous, ongoing star formation 
can (in principle) mask the presence of a more massive old stellar 
population.  The two--component models provided stellar masses that were, 
on average, $\approx 3$ times larger than those from the one--component 
models, with a 68\% upper bound $\approx 6$--8 times more massive.  
For galaxies at lower redshifts, we find that this mass ratio 
between the two-- and one--component models is generally smaller 
for several reasons.  First, at $z < 2$ the infrared photometry reaches 
redder rest--frame wavelengths, and thus provides a better constraint on 
the old stellar content.  Second, many galaxies at $z < 2$ are redder
than the LBGs (evidently with more quiescent star formation relative
to their total stellar masses), and thus cannot hide as much stellar 
mass in a maximally old component.  

\section{Galaxy light and mass in the HDF--N
\label{section:analysis}}

\begin{figure}
\plotone{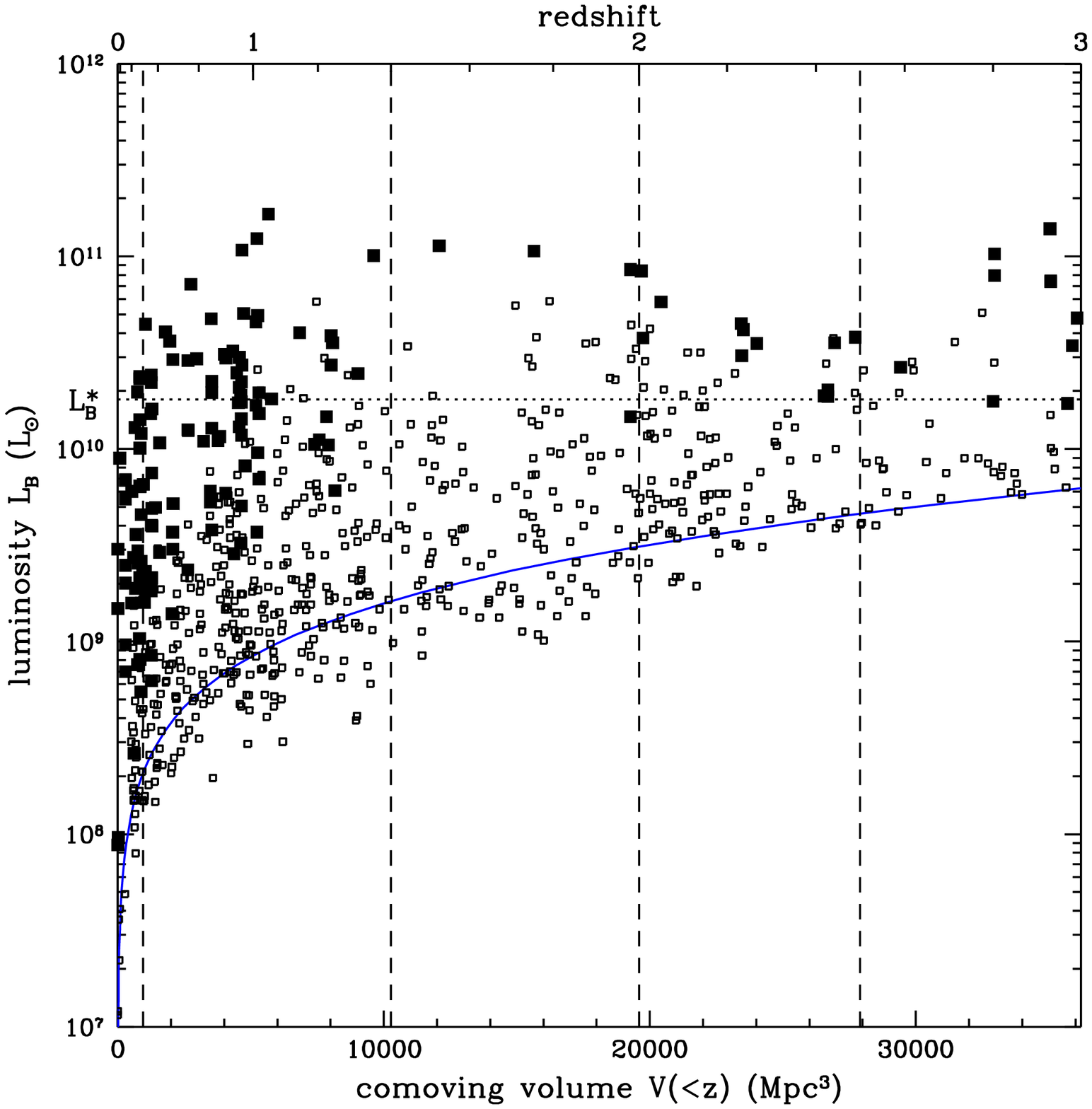}
\caption{
Rest--frame $B$--band luminosities of galaxies in the HDF--N
plotted against redshift $z$ (top axis) or co--moving volume out
to redshift $z$ (bottom axis).  Filled symbols indicate galaxies
with spectroscopic redshifts, while open symbols are galaxies
with photometric redshifts only.  The horizontal dotted line
marks the luminosity of a present--day $L^\ast_B$ galaxy from
the 2dFGRS luminosity function \citep{nor02}.  Vertical dashed lines
divide the four redshift intervals used for analysis in this
paper, each of which encloses approximately $10^4$~Mpc$^{-3}$.
The curve indicates an approximate completeness limit for the
sample for a magnitude limit of $\nich = 26.5$ and assuming
a spectral energy distribution that is constant
in $f_\nu$ flux density.
}
\label{fig:LBvsVz}
\end{figure}

Figure~\ref{fig:LBvsVz} shows the rest-frame $B$--band luminosities
of HDF--N galaxies as a function of redshift $z$ (top axis), and 
of the co--moving volume out to redshift $z$ (bottom axis).  This plotting
scheme has the virtue that the horizontal density of points shows the 
co--moving space density of objects.  The rest--frame $B$--band luminosities 
were computed by interpolating between the fluxes measured in the 
0.3-2.2$\mu$m observed bandpasses.  We converted to solar units assuming
a $B$--band specific luminosity for the Sun of 
$L_{\nu,\odot} (B) = 3.17 \times 10^{18}$~erg~s$^{-1}$~Hz$^{-1}$.
The curve in Figure~\ref{fig:LBvsVz} 
shows an approximate completeness limit for the NICMOS sample, 
assuming a magnitude limit $\nich < 26.5$ and a flat--spectrum $f_\nu$ spectral 
energy distribution.  Roughly speaking, the upper envelope of galaxy 
luminosities is similar for $0 < z < 3$.  The most optically 
luminous galaxy in the HDF--N is a red, giant elliptical FR~I 
radio galaxy at $z = 1.050$ (4-752), although another (possibly
binary) galaxy at $z = 2.931$ (4--52) comes close.

\begin{figure}
\plotone{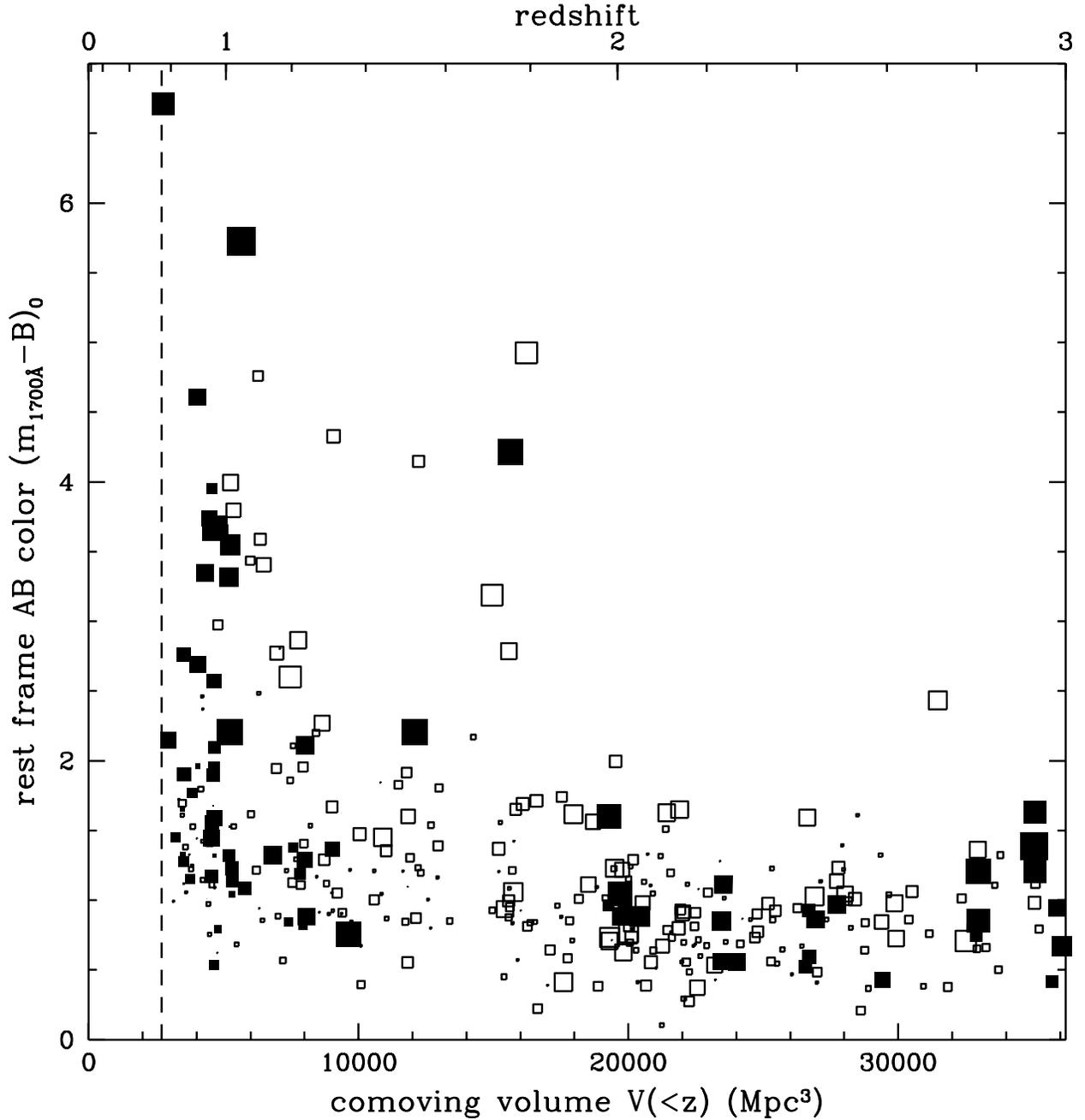}
\caption{
Rest--frame 1700\AA$-B$ colors (AB magnitudes) of HDF--N galaxies 
versus co--moving volume or redshift.  Filled and empty symbols
are as in Figure~\ref{fig:LBvsVz}, and the symbol size scales
with the rest--frame $B$--band galaxy luminosity.  No direct
measure of the 1700\AA\ rest--frame light is available for
galaxies at $z < 0.75$ (vertical dashed line).  The galaxy at 
$z \approx 2.7$ which is unusually red (1700\AA$-B \approx 2.5$) 
for its redshift is HDFN~J123651.7+621221, which exhibits
hard X--ray \citep{hor00} and radio \citep{ric99} emission,
and is suspected of being a type~II AGN.   It may also be marginally 
detected at 15~$\mu$m \citep{aus99} and 1.3~mm \citep{dow99} as well.
}
\label{fig:UV17mBvsVz}
\end{figure}

Although bright galaxies have similar space densities in the HDF--N 
out to $z = 3$, their stellar populations do not remain the same.  
Figure~\ref{fig:UV17mBvsVz} presents the rest--frame UV--optical 
colors of HDF--N galaxies versus redshift, showing a strong trend 
toward bluer colors at higher redshift.  This effect cannot be 
attributed to selection biases because the sample is selected at 
near--infrared wavelengths.  All points shown in Figure~\ref{fig:UV17mBvsVz} 
are measurements, not limits;  no object from the NICMOS sample 
studied here is undetected in the WFPC2 HDF--N.  For galaxies
at $z \sim 2.7$, the rest--frame 1700--$B$ color is approximately 
matched by the observed $V_{606}-H_{160}$.  For the range of galaxy 
magnitudes we find at that redshift ($23 \lesssim H_{160} < 26.5$), 
we could have measured colors as red as 
$6 \gtrsim V_{606} - H_{160} > 4$, but none are found.  The trend toward 
bluer colors at higher redshift presumably reflects a tendency toward 
younger stellar populations and increasing levels of star formation in 
more distant galaxies.  As noted by PDF, nearly all HDF--N galaxies
at $z > 2$ have colors much bluer than those of present--day
Hubble sequence galaxies, more similar to those of very active, 
UV--bright starburst galaxies.  We may thus expect that the typical 
stellar mass--to--light ratios are lower for galaxies at higher redshift.

\begin{figure}
\plotone{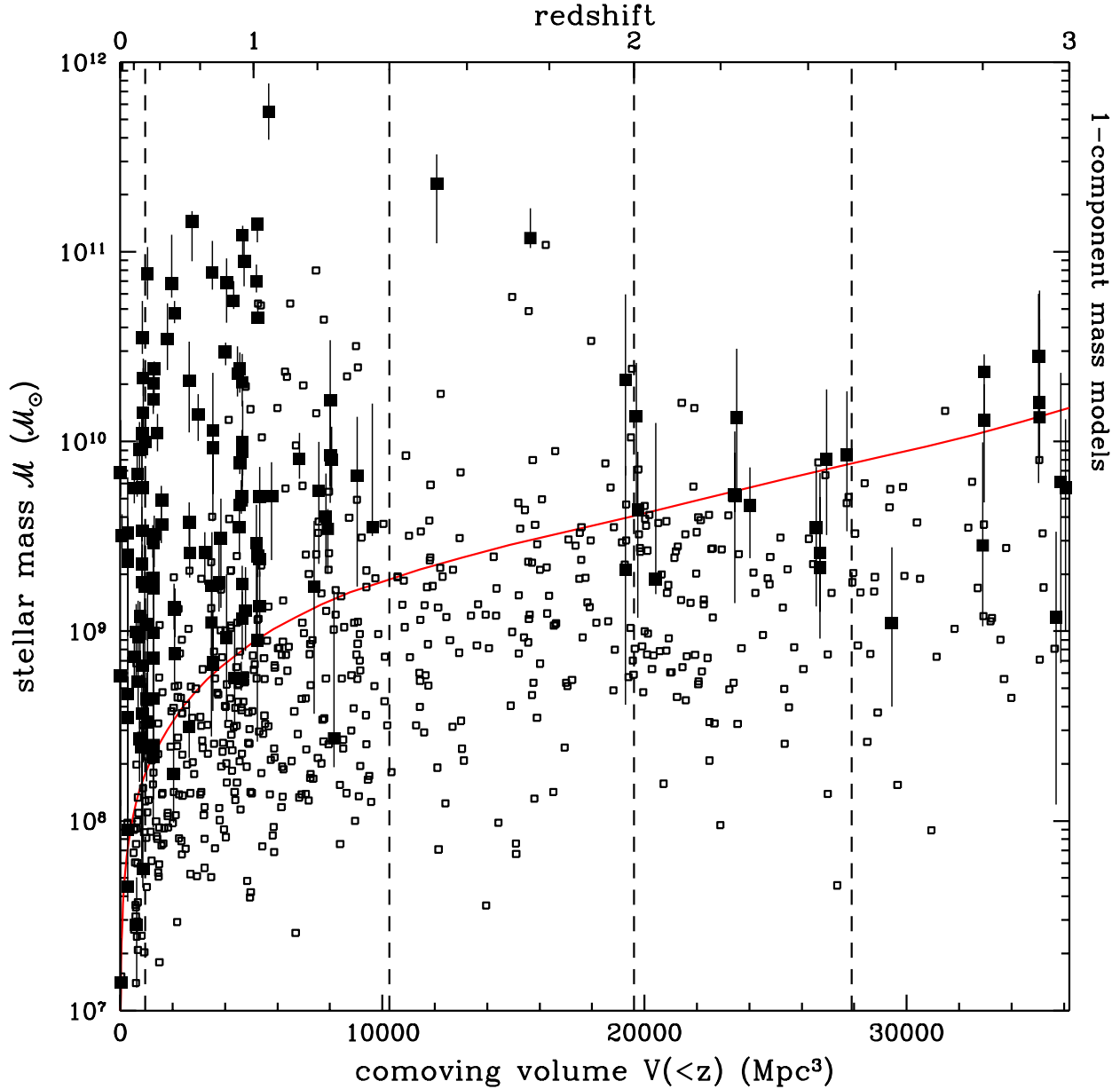}
\caption{
Estimates of stellar mass for HDF--N galaxies versus co--moving 
volume or redshift.  Symbols are as in Figure~\ref{fig:LBvsVz}.
The mass estimates use Bruzual \& Charlot models with solar metallicity
and the 1--component star formation histories 
($\Psi(t) \sim \exp(-t/\tau)$) described in the text.  Error bars
showing 68\% confidence ranges are shown for the spectroscopic
sample only in order to avoid overcrowding the figure.  Masses 
estimated from lower metallicity models are generally somewhat smaller, 
while the ``maximum $\mathM/L$'' 2--component star formation histories 
yield larger masses, especially for the objects at higher redshift.
The general character of the diagram remains the same, however,
whichever models are used.  The curve shows the mass corresponding 
to a maximally old stellar population, formed in a burst at $z = \infty$ 
and aging passively, with $\nich = 26.5$.  The NICMOS sample should
be complete for all galaxies with stellar masses above this line.
Galaxies younger and bluer than this maximally old model can be
detected to lower mass limits.
}
\label{fig:Mass1vsVz}
\end{figure}

Figure~\ref{fig:Mass1vsVz} shows the best--fitting stellar mass 
estimates for HDF--N galaxies, calculated using the 1--component
(i.e., single exponential) SFR history models described in 
\S\ref{section:masses}.  Unlike the luminosities plotted in 
Figure~\ref{fig:LBvsVz}, the upper envelope of galaxy masses declines 
markedly at higher redshift, and the space density of massive 
galaxies (say, $\mathM > 10^{10}\mathM_\odot$) thins out.   
PDF found that the characteristic stellar mass of HDF--N LBGs 
is $\sim 10^{10}\mathM_\odot$.  Even using a complete, 
infrared--selected, photometric redshift sample (which 
should be free of rest--frame UV selection biases), we do 
not find additional galaxies at similarly large redshifts
with stellar masses greater than those of the bright HDF--N
LBGs.   The curve in the figure shows the stellar mass 
corresponding to a maximally old, red stellar population 
formed at $z = \infty$, aging passively, with $\nich = 26.5$.
The NICMOS sample should be complete for galaxies with 
$\mathM > 10^{10} \mathM_\odot$ nearly out to $z = 3$.  
At any redshift, some galaxies below this curve -- those objects 
that are bluer than the maximum--$\mathM/L$, passive evolution 
limit -- can be detected.   However, we cannot rule out the 
possibility that we are missing faint, red, fading galaxies
with masses that fall below the curve in Figure~\ref{fig:Mass1vsVz},
but which might be too faint to detect in the NICMOS images.  

As discussed in \S\ref{section:masses}, we may set an upper bound
on the stellar masses by adopting 2--component SFR models which include
a maximally old component formed at $z = \infty$.  This has a greater 
effect on the masses of the galaxies at higher redshift, where 
the observed colors are generally bluer and the photometric measurements
do not reach as far to the red in the galaxy rest frame.  Nevertheless,
the basic character of Figure~\ref{fig:Mass1vsVz} remains the same
even when these ``maximal'' mass estimates are used.  Even with
the maximum $\mathM/L$ models, very few galaxies at $z > 1.4$,
and no galaxies at $2 < z < 3$, have $\mathM > 10^{11}\mathM_\odot$.
Using either SFR model, the most massive galaxy in the HDF--N
is the $z = 1.05$ gE 4-752.0, and the most distant galaxy with
$\mathM > 10^{11}\mathM_\odot$ is the gE host galaxy (4-403.0)
of the $z \approx 1.75$ type Ia supernova SN1997ff \citep{gil99,rie01}.

We will not discuss galaxies at $z > 3$ in this paper because 
the NICMOS data do not probe the rest--frame optical light at 
those redshifts.  However, one object which warrants special 
attention is the ``$J$--dropout'' object J123656.3+621322
\citep{dic00a}, which has the reddest near--infrared colors
of any object in the HDF--N.  Its redshift is unknown;
it might conceivably be an LBG or QSO at $z > 12$, a dusty object 
at intermediate redshift, or an evolved galaxy with an old 
stellar population at high redshift.  This latter possibility 
is most relevant here, as it might imply a large stellar 
$\mathM/L$.   If we adopt a best--fit $z_{phot} = 3.3$ for 
this object (excluding the $z > 12$ option), we find that this 
object can be fit with an old ($\sim 1.6$~Gyr) stellar population 
with moderate dust extinction (required to match the red $J-H$ 
color -- cf.~\citet{dic00a}), and $\mathM \approx 1\times10^{11} \mathM_\odot$.  
If this interpretation were correct, this object would be 
significantly more massive than any other HDF--N galaxy at $z > 2$.
The only other object at $z > 3$ with a fitted $\mathM > 10^{11}\mathM_\odot$
is a broad--line AGN (4-852.12, $z = 3.479$), whose luminosity and colors 
are probably significantly affected by the active nucleus, and where 
stellar population model fitting is therefore inappropriate.  
The comparatively red (and X--ray luminous) galaxy HDFN~J123651.7+621221, 
at $z_{phot} = 2.7$ (see Figure~\ref{fig:UV17mBvsVz}),
is fit with a modest $\mathM \approx 1.5\times10^{10} \mathM_\odot$.

\section{The global luminosity and stellar mass densities
\label{section:results}}

The HDF--N is a very narrow pencil--beam survey along a single sight--line,
and any statistical analysis of its most luminous or massive galaxies is 
subject to small number statistics and to density variations due 
to clustering.   We therefore adopt a very crude approach here,
dividing the volume at $0.5 < z < 3$ into four redshift bins
with approximately equal co--moving volume $\approx 10^4$~Mpc$^3$ each:
$z_l$ to $z_u = 0.5$--1.4, 1.4--2.0, 2.0--2.5, and 2.5--3.0.  The 
co--moving volume at $0 < z < 0.5$ is $< 1000$~Mpc$^3$, and we do not 
consider galaxies in that redshift range further here, although 
nearly 40\% of the lifetime of the universe elapses in that interval.
Based on the 2dFGRS optical luminosity function of \citet{nor02},
in the local universe we would expect $\sim 12$ galaxies
with $L \geq L_B^\ast$ in a volume of $10^4$~Mpc$^3$.
From the 2dF+2MASS stellar mass function of \citet{col01},
we would expect $\sim 6$ galaxies with 
$\mathM \geq \mathM^\ast (= 1.4\times 10^{11} \mathM_\odot)$ 
in the same volume.

\begin{figure}
\plotone{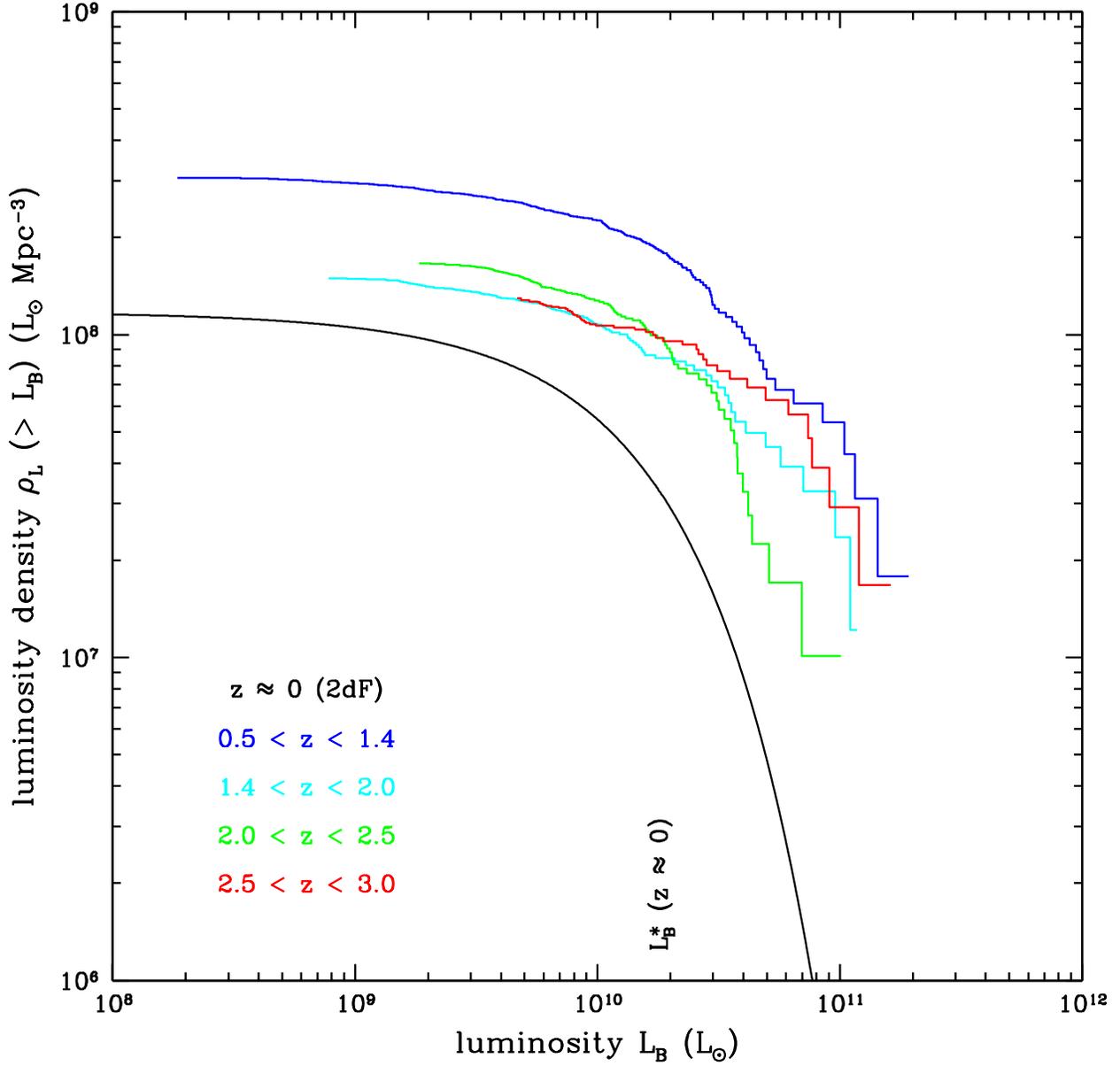}
\caption{
The rest--frame $B$--band luminosity density contained in galaxies
brighter than a given luminosity, plotted for various redshift intervals.
The lowest curve is an integral over the 2dFGRS luminosity function,
while the upper histograms show HDF--N galaxies in the redshift
intervals marked in Figure~\ref{fig:LBvsVz}.
}
\label{fig:rhoL}
\end{figure}

Figure~\ref{fig:rhoL} shows the cumulative luminosity densities 
$\rho_L(> L_B)$ from the local 2dF luminosity function and from 
the HDF--N redshift slices.  The HDF--N data are deep enough so that 
the luminosity densities are roughly convergent within each redshift slice
at the detection limit.  In each redshift interval, we evaluate the 
integrated $B$--band luminosity density and its uncertainty in several ways.  
First, we obtain a direct estimate, without extrapolation to fainter 
luminosities, by summing the galaxy luminosities, weighted inversely
by the volumes over which they could be detected (the $1/V_{max}$
formalism of \citet{sch68}).  For each galaxy, the maximum redshift 
to which it could be seen with $\nich < 26.5$ is calculated, 
using the multiband photometry for each object to compute the 
$k$--corrections directly.  Most galaxies are visible throughout their 
redshift slices $z_l < z < z_u$, but when $z_{max} < z_u$, then $z_{max}$ 
is used to compute the $1/V_{max}$ weighting.   The volume--weighted 
luminosities are then summed, and the uncertainty in this value is 
estimated by bootstrap resampling.  Second, we have fit \citet{sch76} 
functions to the number densities in $\log(L_B)$ bins, estimating 
parameter uncertainties using Monte Carlo simulations where the numbers 
of objects per bin were varied within a Poisson distribution.  
The Schechter functions generally give acceptable fits, except 
for the lowest redshift interval, where reduced $\chi^2 = 2.1$, mainly
due to an inflection at the faint end which has little impact on the 
integrated luminosity density.\footnote{Note that the universe nearly doubles 
in age between $z = 1.4$ and 0.5.  The faintest galaxies used in the
luminosity function analysis are all at the low end of the redshift 
range.  It may not be surprising that a simple Schechter function does 
not fully represent the (quite likely evolving) luminosities in this interval.}
The individual Schechter function parameters, $\phi^\ast$, $L^\ast$, 
and $\alpha$, are moderately well constrained for the lowest redshift 
subsample ($0.5 < z < 1.4$), where 382 objects are used, spanning nearly 
a factor of 1000 in luminosity.  The best--fitting Schechter function 
parameters for this redshift bin, expressed in the $b_J$ band 
(conventionally normalized to Vega) for comparison to local determinations,
are $M^\ast_{b_J} = -20.47 \pm 0.35 + 5\log h$,
$\phi^\ast = (2.0 \pm 0.5) \times 10^{-2} h^3$~Mpc$^{-3}$, and 
$\alpha = -1.16 \pm 0.07$.
The luminosity function parameters are more poorly determined at 
higher redshifts, but in general the integrated luminosity densities, 
$\rho_L = L^\ast \phi^\ast \Gamma(\alpha+2)$, are better constrained.

Table~\ref{tab:lumdenstab} presents the results of the luminosity density
integration using both methods.  At $z < 2.5$, the Schechter function
integrations yield values for $\rho_L$ which are within 0.1~dex of the 
direct sums, confirming that the HDF observations are detecting most of 
the light at those redshifts.  In the highest redshift interval ($2.5 < z < 3$), 
our sample has only 45 galaxies spanning a range of $\sim 30$ in luminosity, 
and the Schechter function slope $\alpha$ is poorly constrained,
leading to larger uncertainty in $\rho_L$;  $>5$\% of the Monte Carlo
simulations lead to divergent $\rho_L$ (i.e., $\alpha < -2$).
For the purposes of this analysis, we have refit the data in the highest
redshift bin fixing the slope to $\alpha = -1.4$, the steepest value
found in the lower redshift ranges, but still somewhat shallower than 
$\alpha = -1.6$ for the rest--frame UV luminosity function 
of LBGs \citep{ste99}.

\begin{figure}
\plotone{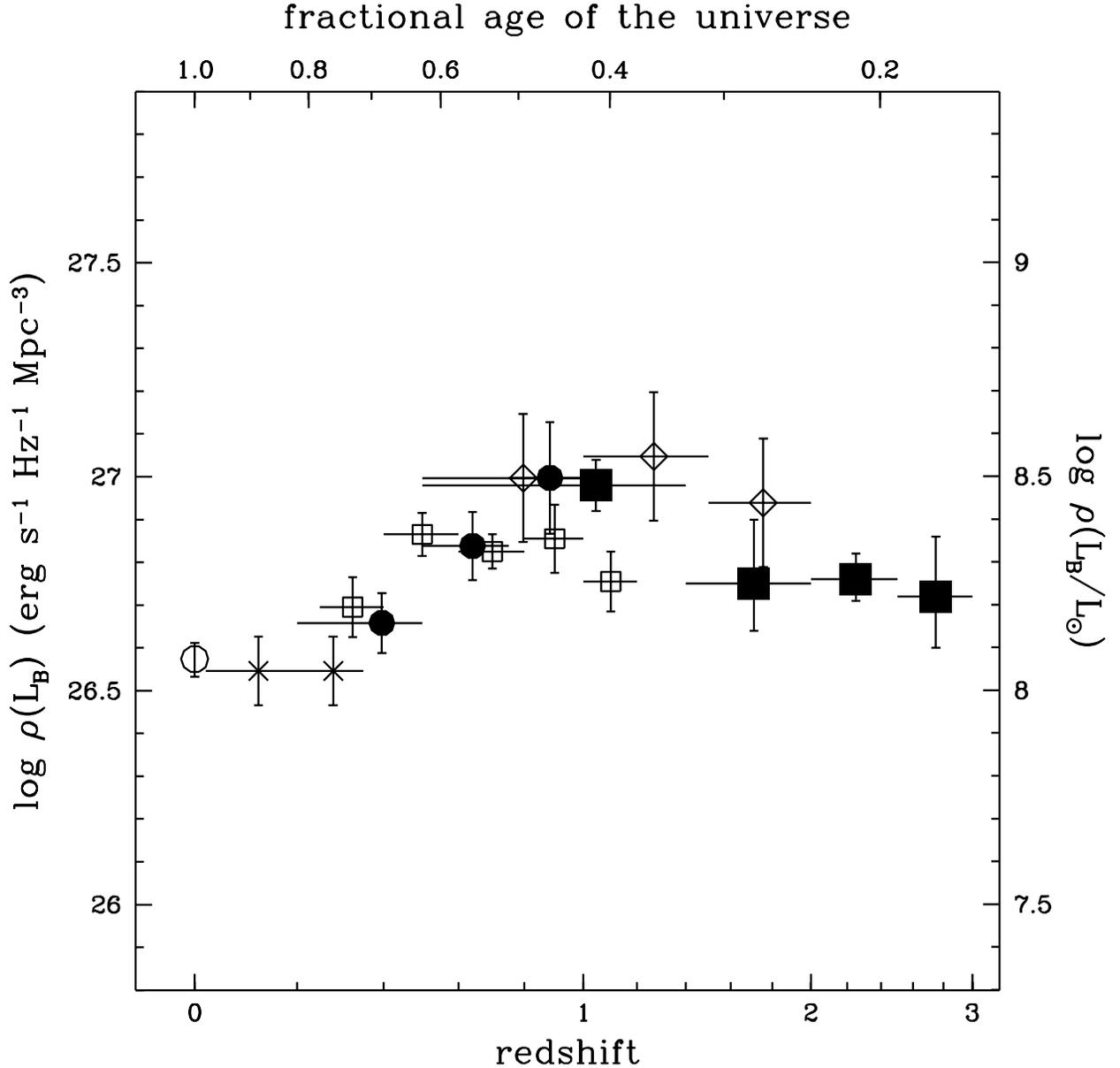}
\caption{
The co--moving rest--frame $B$--band luminosity density as a function
of redshift. The horizontal axis scaling is linear in $\log(1+z)$.
Our HDF--N results are shown as large filled squares.
For each point, the horizontal bar shows the redshift range of
each bin, while the point position is the volume midpoint within 
this range.   Other points show values from the literature, 
rescaled to our adopted cosmology 
(open circle -- \citet{nor02};
crosses -- \citet{ell96};
filled circles -- \citet{lil96};
open squares -- \citet{wol02};
diamonds -- \citet{con97}, tabulated in \cite{mad98}).
The local value from \citet{nor02} includes an evolutionary 
correction and is therefore plotted at $z = 0$;  the measurements
come from galaxies at $0 < z \lesssim 0.25$.  Values from 
\citet{con97} are also based on HDF--N data, using an earlier 
photometric redshift analysis without NICMOS data.  
}
\label{fig:rhoLvsz}
\end{figure}

Our estimates of the rest--frame $B$--band luminosity density,
along with various measurements from the literature, 
are shown in Figure~\ref{fig:rhoLvsz}.
At $z \approx 1$, we find that $\rho(L_B)$ is
$\sim 2.5\times$ larger than that in the local universe 
($\log (\rho_L/L_\odot) = 8.1$, \citet{nor02}), consistent
with previous results.\footnote{
At $0.75 < z < 1.0$, \citet{lil96} find 
$\log (\rho_L/L_\odot) = 8.50 \pm 0.13$, where we have adjusted
their measurement to the cosmology adopted here.
This value is in excellent agreement with our value at 
$\langle z \rangle = 1.05$ ($0.5 < z < 1.4$).   Recent
results based on photometric redshifts from the COMBO--17
survey find a somewhat smaller value, 
$\log (\rho_L/L_\odot) = 8.35 \pm 0.08$, at $0.8 < z < 1$ 
\citep{wol02}.}
The luminosity function and density are consistent with
modest luminosity evolution (0.85~mag for rest--frame $L^\ast_B$)
and 25\% higher space densities ($\phi^\ast$) at $z \approx 1$ 
compared to the present day.  At higher redshift ($1.4 < z < 3$),
the luminosity density is roughly constant at $\sim 1.5\times$ the 
local value.  However, we have seen that galaxies at higher 
redshift have bluer colors and presumably lower mass-to-light
ratios.  Hence, the density of stellar {\it mass} may be expected 
to evolve differently than that of the light.

To compute the stellar mass density in each redshift slice,
we begin by summing the masses for everything that we can see,
e.g., $\rho_\ast (> L_B)$.  As noted above, for the oldest 
possible stellar populations with maximum $\mathM/L$ we can only 
be certain of being complete down to the mass limits shown by the 
line in Figure~\ref{fig:Mass1vsVz}.  Thus we can confidently
detect objects with $\mathM > 4\times10^{9} \mathM_\odot$ out to
$z = 2$ and $\mathM > 10^{10} \mathM_\odot$ out to nearly $z = 3$, 
but we cannot be completely certain that some objects with smaller
masses but large $\mathM / L$ are not missed.  As noted above,
there is little evidence for a significant population
of galaxies at $z > 2$ with $\mathM / L$ much larger than that seen 
for LBGs, so it might be surprising if such a population were to
appear fainter than the detection limit of the NICMOS data.
Also, at all redshifts in the HDF--N sample, the mean 
$\langle \mathM/L \rangle$ is either declining or flat with
decreasing $L_B$ (cf.\ Figure~1 from PDF, where a mild tendency
toward bluer colors at fainter magnitudes is seen for LBGs at
$2 < z < 3.5$).   This is a well--known trend in the local universe
(cf.\ \citet{kau02}):  fainter galaxies on average tend to have bluer 
colors and lower stellar $\mathM/L$.   This effect would be even more 
pronounced if our stellar population modeling had incorporated a trend 
of increasing metallicity with mass or luminosity.

Parenthetically, it is worth noting that \citet{sha01} have reported 
evidence for an extremely steep faint--end slope to the rest--frame 
optical luminosity function of LBGs at $z \approx 3$, as well as a tendency 
for fainter galaxies to have redder UV--optical colors, and hence presumably 
larger $\mathM/L$.  These two trends are intimately connected in the 
\citet{sha01} sample, since they infer the observed--frame $K_s$--band 
luminosity distribution from the observed ${\cal R}$--band data 
(with which \citet{ste99} measured $\alpha_{UV} = -1.6$) and the 
${\cal R} - K_s$ color distribution of galaxies in their sample.  
Taken at face value, these facts would imply a very large mass density 
of faint red galaxies at $z \approx 3$, below the magnitude limits 
of ground--based Lyman break galaxy surveys.   The \citet{sha01} sample, 
however, spans a limited range of luminosities, restricted to the bright end 
of the LBG distribution.  It is very difficult to constrain the faint 
end slope with such data;  we are unable to do so reliably even with our
deeper NICMOS sample.  Moreover, we find no such reddening trend with 
fainter magnitudes for HDF--N LBGs, pushing well below the magnitude 
limits of the ground--based samples.  If anything, the opposite seems 
to be the case.  For the moment, therefore, we discount this result 
from \citet{sha01}, but we stress the important 
caveat that even the deepest near--infrared data and photometric redshift 
samples presently available cannot definitely constrain these important 
properties of the galaxy distribution at $z > 2.5$, leaving genuine 
uncertainty about the integrated stellar mass content at such redshifts.

\begin{figure}
\plotone{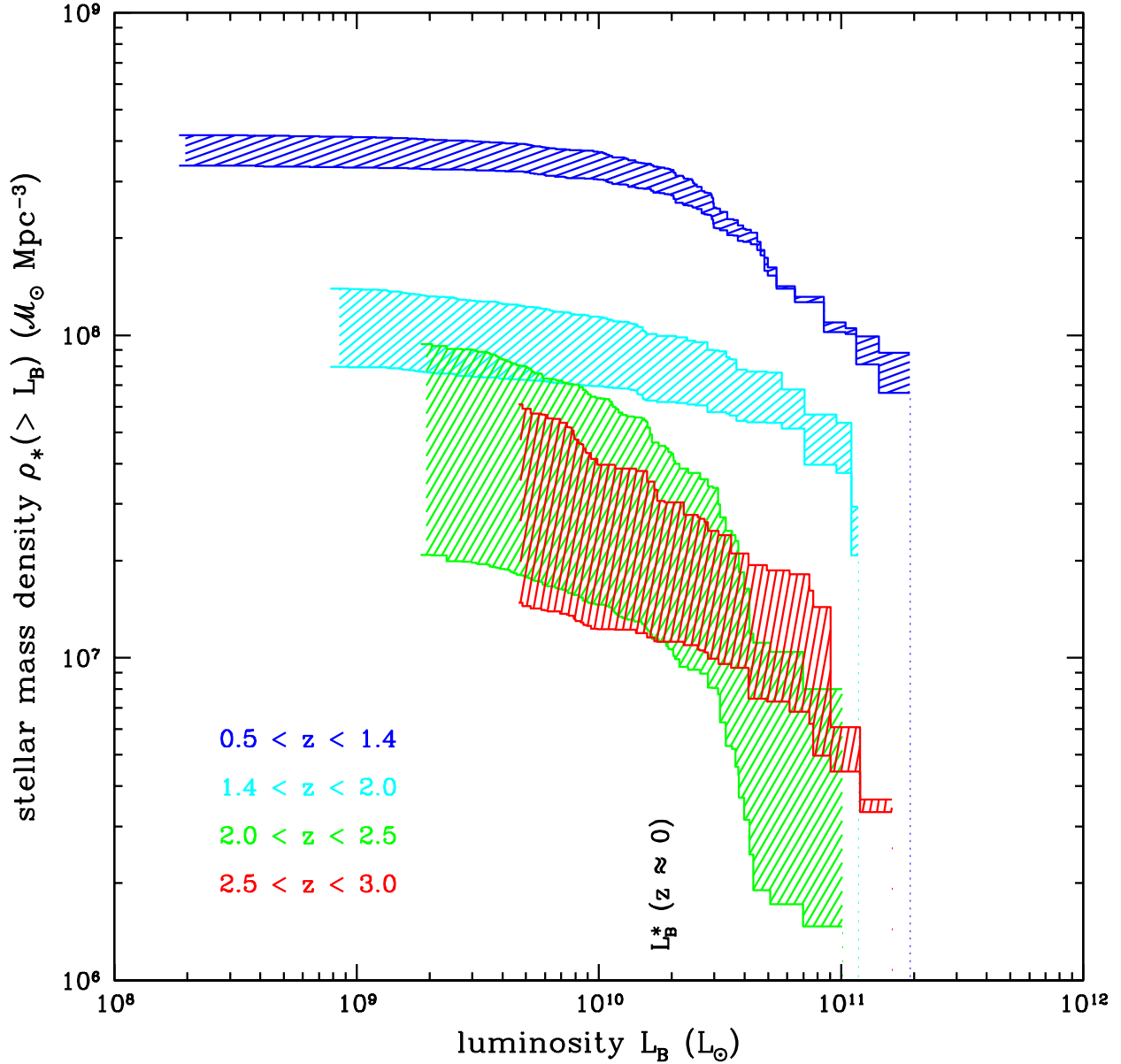}
\caption{
Cumulative stellar mass densities for HDF--N galaxies brighter
than a given rest--frame $B$--band luminosity, in various
redshift intervals.  The hatched regions show the range of
estimates using different stellar population models, going from
low metallicity, 1--component $\Psi(t)$ models (lower bounds)
to solar metallicity, 2--component ``maximum $\mathM/L$'' models
(upper bounds).
}
\label{fig:rhoM_all}
\end{figure}

The cumulative stellar mass density $\rho_\ast(> L_B)$ in the
four HDF--N redshift slices is presented in Figure~\ref{fig:rhoM_all}.
As discussed above, the fitted mass estimates depend on 
the model metallicities and on the adopted star formation histories.  
Here, we show the range of mass densities obtained by varying the model
metallicities from 0.2 to $1.0 \times Z_\odot$, and from the
single and 2--component SFR history models.   The dependence of
the mass densities on these model assumptions
is less than a factor of 2 at $z \sim 1$, but can be as much as 
a factor of 4 out at $z \sim 2.7$.   At $z < 2.5$, the mass densities
are approximately convergent.  At $z > 2.5$, $\rho_\ast$ from the 
1--component models appears to be converging, but is still
rising for the 2--component, maximum $\mathM/L$ models.
This is a consequence of the mass completeness limit drawn
in Figure~\ref{fig:Mass1vsVz}.  Just as galaxies might be present,
but undetected by NICMOS, with masses that fall below that curve,
any faint galaxy that {\it is} detected may in principle 
have an unseen old stellar component with mass roughly equal to 
that shown by the completeness line.  The maximum $\mathM/L$,
2--component models therefore permit the potential existence of 
a galaxy population with $\mathM / L \propto L_B^{-1}$ below 
some threshold mass set by the NICMOS detection limits.

For each redshift interval, we calculate the volume--averaged
mass--to--light ratio $\langle \mathM/L \rangle$ by taking the ratio 
of the mass and $B$--band luminosity densities for the observed 
galaxy populations.  Our calculations of $\langle \mathM/L \rangle$
under various stellar population assumptions are presented in
Table~\ref{tab:moltab}.  For comparison, the 2dFGRS measurements
of \citet{nor02} and \citet{col01} yield a local value 
$\langle \mathM/L \rangle = 4.4 \mathM_\odot / L_\odot$.
This is significantly larger than the values we derive at high 
redshift, implying substantial evolution in the mean stellar 
population properties (averaged over all galaxies) from 
$z \approx 1$ to the present.  We multiply the computed 
$\langle \mathM / L \rangle$ values by our estimates of the total 
$\rho_L$ from the Schechter function integration described above.  
Note that if $\mathM/L$ decreases at fainter luminosities, this 
procedure will slightly overestimate the stellar mass density.
The random uncertainty in $\rho_\ast$ is 
estimated from the combined uncertainties in $\rho_L$ and the quadratic sum 
of the mass fitting uncertainties (for each particular choice of model 
metallicity and $\Psi(t)$) for the galaxies in each redshift range.
The systematic dependence of the results on these choice of stellar
population model assumptions is illustrated below, and our 
results are tabulated in Table~\ref{tab:rhoMtab}.
Note that we have used the rest--frame $B$--band primarily as an 
accounting device here, calculating $\langle \mathM/L_B \rangle$ 
and using a Schechter fit to the luminosity function to correct for 
sample incompleteness at the faint end.   Our stellar masses are 
actually derived using NICMOS and ground--based near--IR observations 
that reach significantly redder rest--frame wavelengths (0.54--1.44$\mu$m
for the $K_s$--band data) over the entire redshift range being 
considered here.  

Regardless of the model adopted for the mass estimates,
there is good evidence that the global stellar mass density
increased with cosmic time.   E.g., for galaxies
brighter than present--day $L^\ast_B$, which are abundant 
in the HDF--N at all redshifts $z < 3$, the minimum 
estimate for $\rho_\ast$ at $z \approx 1$ is $\sim 7\times$ 
larger than that from the maximum $\mathM/L$ models at $z \approx 2.7$.   

Figure~\ref{fig:OmegaStar_a} places the HDF--N mass density estimates 
into a global context, comparing them with other data at $0 < z < 1$, 
where the HDF--N itself offers little leverage due to its small volume.  
The local stellar mass density is taken from \citet{col01}, 
integrating over their fit to the stellar mass distribution of 
2dFGRS galaxies.  This value is consistent with earlier estimates
by \citet{fuk98} and \citet{sal99}, but has smaller statistical
uncertainties.  Values at $0.3 < z < 1$ are taken from \citet{bri00} 
and \citet{coh02}, with adjustment for the change in cosmological 
parameters to those adopted here.  Our own estimates of $\Omega_\ast$, 
as well as those of \citet{col01} and \citet{coh02}, integrate
over luminosity functions to estimate the total luminosity and mass 
densities.  \citet{bri00} computed $\rho_\ast$ by summing over 
galaxies in the range $10.5 < \log (\mathM/\mathM_\odot) < 11.6$, 
without extrapolating further down an assumed mass function, and thus
their published values are strictly lower limits.  Those authors 
estimate that their summation accounts for $\geq 80$\% of $\rho_\ast$, 
and we have therefore applied a 20\% upward correction to their points to roughly 
account for this incompleteness.  Note also that \citet{col01} fit metallicities
for each galaxy, primarily based on near--infrared colors.   From their 
Figure~11, it appears that most 2dFGRS galaxies are best fit with models 
close to solar metallicity, and generally not lower than $Z = 0.004$.  
\citet{bri00} also allow metallicity to vary
from 0.004 to 0.02 in their mass fitting, while \citet{coh02} adopt
evolutionary corrections based on population synthesis models from 
\citet{pog97}, which include chemical evolution.   We have found 
(see PDF) that we cannot set reliable photometric constraints on metallicity 
for the higher redshift galaxies, and therefore have used different (but fixed)
values to explore the systematic dependence of $\rho_\ast$ on $Z$.

\section{Discussion
\label{section:discussion}}

\subsection{The evolution of the global stellar mass density
\label{section:discuss_evol}}

At $z \approx 1$, we find that 50--75\% of the present--day stellar mass 
density had already formed, depending on the stellar population models adopted.
Our estimates are consistent with those from \citet{bri00} and \citet{coh02},
who studied larger samples over greater volumes at $0.3 < z < 1$.  Our HDF--N 
analysis alone does not offer sufficiently fine redshift resolution to 
further pin down the epoch when (e.g.) 50\% of the stars in the universe 
had formed, but it seems to have taken place somewhere in this range.  

At $z > 1.4$, we see a steep drop in the total stellar mass density with 
increasing redshift.  The systematic uncertainties in our mass estimates 
are large at $z > 2$, primarily because the NICMOS near--infrared measurements 
sample only rest--frame optical wavelengths in the $B$-- to $V$--band range, 
loosening our constraints on $\mathM/L$.  In addition, even these NICMOS 
observations do not go deep enough to tightly constrain the luminosity function 
at $z > 2.5$, although a fairly robust estimate of the luminosity density 
can be made with only modest and reasonable constraints on the faint 
end slope $\alpha$.  Nevertheless, the apparent change in the stellar mass 
density is sufficiently dramatic as to exceed the substantial uncertainties 
in the measurements.  For our fiducial stellar population models with solar 
metallicity and exponential star formation histories, the mass density 
changes by a factor of 17 from $z \approx 2.7$ to the present--day.  
Even for the ``maximum $\mathM/L$'' 2--component models, $\rho_\ast$ 
is $\sim 7\times$ smaller at $z \approx 2.7$ than at $z = 0$.  Even if all 
galaxies in the highest redshift bins were set at their 68\% upper confidence 
limits of their mass estimates for this model, we would find less than 
1/3rd of the present--day mass density present.   

\begin{figure}
\epsscale{0.9}
\plotone{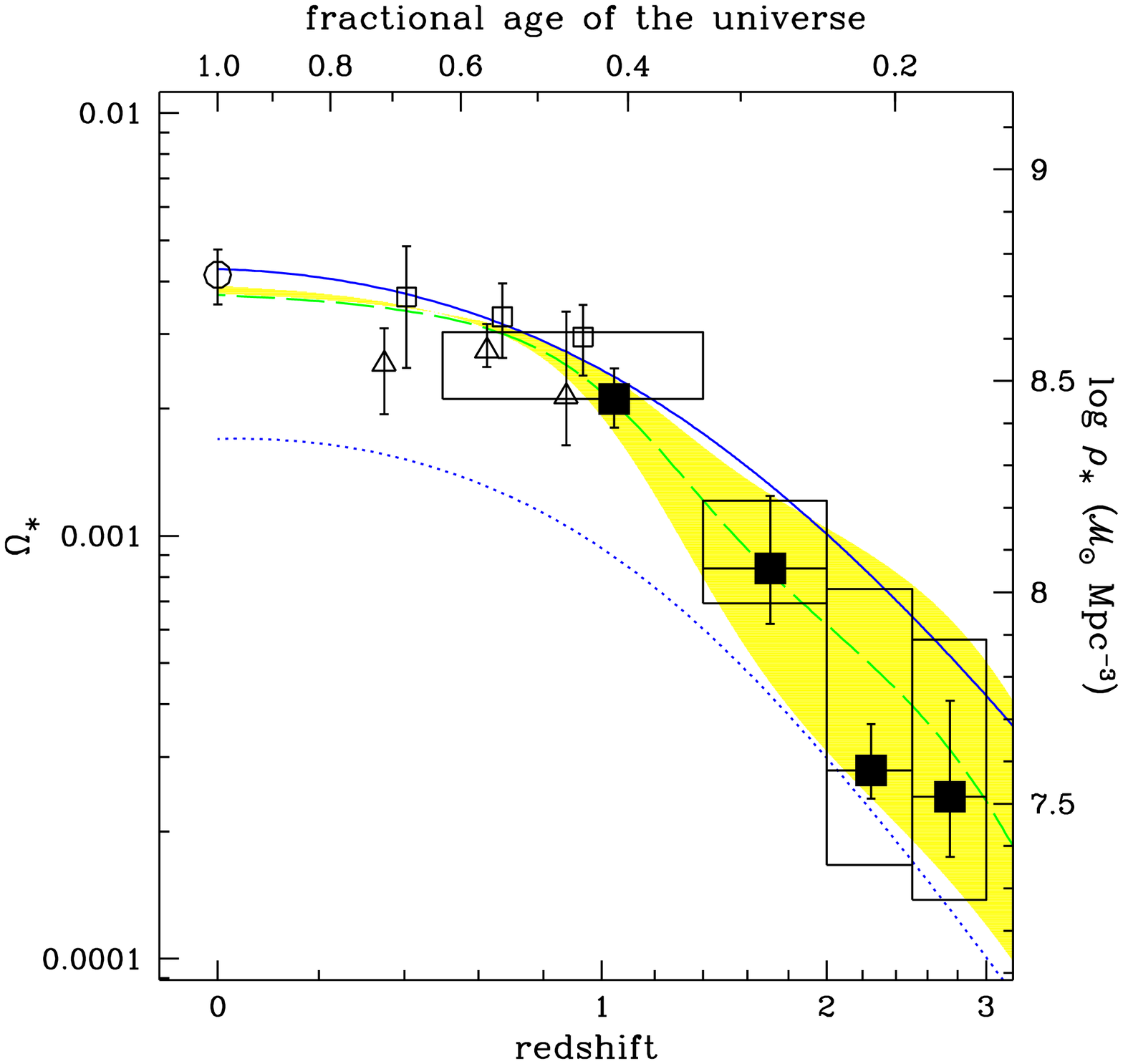}
\caption{
The redshift evolution of the co--moving stellar mass 
density, $\Omega_\ast$.   The horizontal axis scaling and error bars 
are as in Figure~\ref{fig:rhoLvsz}.
Open symbols show results from the literature at $0 < z < 1$ 
(circle -- \citet{col01}; triangles -- \citet{bri00}; 
squares -- \citet{coh02}).  
The HDF--N points (large filled squares) show results for the $Z_\odot$, 
1--component $\Psi(t)$ model fits.  The vertical error bars show the 
68\% random errors for this choice of model, combining the sampling 
variance and the mass fitting uncertainties.  The vertical extent of 
the boxes show the range of systematic uncertainty introduced by 
varying the metallicity and the star formation histories of the 
mass--fitting model.  The blue curves show the result of integrating 
the cosmic star formation history SFR($z$) traced by rest--frame UV 
light, with and without corrections for dust extinction (solid and 
dotted curves).  The dashed green curve shows the integrated star
formation history from \citet{pei99}, with their 95\% confidence 
range indicated by the shaded region.
}
\label{fig:OmegaStar_a}
\end{figure}

The quantity $\rho_\ast (z)$ is essentially the time integral over 
the cosmic star formation history 
SFR($z$)~=~$\dot{\rho}_{\ast}(z)$.\footnote{
Note that in our analysis we have not imposed any continuity 
between our estimates of $\rho_\ast(z)$ and the ensemble 
star formation history, $\dot{\rho}_{\ast}(z)$, from the stellar 
population models that were used to derive the masses.  In principle, 
$\rho_\ast$ could even decrease with time, which would be inconsistent 
with any SFR($z$).  (In practice this is not observed.)   
Our ``maximum $\mathM/L$'' models may be logically inconsistent 
with evidence for $\Omega_\ast$ rising with time, since they 
assume that a large fraction (generally the majority) 
of the stellar mass in galaxies formed at infinitely high redshift 
(and thus was always present), whereas even the maximum mass models 
show $\rho_\ast$ rising with time.  These models should therefore be 
regarded as illustrative (at a given redshift) only, representing 
the most conservative assumption we can make without postulating 
large amounts of stellar mass in galaxies that are invisible to 
our NICMOS imaging.
}
The blue curves in Figure~\ref{fig:OmegaStar_a} show the integral 
over SFR($z$) inferred from rest--frame UV light (e.g., \citet{lil96,mad98,ste99}),
using parameterized fits given by \citet{col01} (extrapolated to $z_{max} = 7$).
The integration incorporates the time--dependent recycling of stellar material 
back to the ISM from winds and supernovae, following the relation for a Salpeter 
IMF and solar metallicity calculated with the Bruzual \& Charlot models.  
Asymptotically, this recycled fraction $R = 0.32$ at age~=~13~Gyr.

It is now widely believed that measurements based on rest--frame UV light 
underestimate the total rate of high redshift star formation unless corrections 
for dust extinction are included.  This can also be seen from the integrated 
stellar mass, $\Omega_\ast(z)$, as well.  The curves in 
Figure~\ref{fig:OmegaStar_a}
show the integrated SFR($z$) with and without nominal corrections for dust 
extinction taken from \citet{ste99}.  As \citet{mad98} and \citet{col01} have 
previously noted, without the boost from an extinction correction, this integral
results in a present--day mass density which is substantially smaller than the 
observational estimates.   Here, we can see that this is true over the 
broader redshift range $0 < z < 3$, although at $z > 2$ the inconsistency 
is small given the systematic uncertainties in $\rho_\ast$.  
On the other hand, the extinction--corrected SFR($z$) results in 
$\rho_\ast(z)$ which agrees fairly well with observations at 
$z \approx 0$ and $z \approx 1$,  but which exceeds our best estimates 
at higher redshifts.   Again, given the modeling uncertainties, these 
estimates are consistent with the upper bounds from the maximum 
$\mathM/L$ models, but as noted above, this class of mass models is 
somewhat unrealistic and inconsistent when considered over a wide range 
of redshifts.  The parameterized SFR($z$) from \citet{col01} rises by
a factor of $\approx 2$ from $z = 4$ to its peak at $z \approx 1.5$.
A flatter star formation history at $z > 2$ would exacerbate this 
discrepancy.

\citet{pei99} have used measurements of the UV luminosity density
and neutral gas density as a function of redshift and the far--infrared 
background to constrain a self--consistent solution for the global
star formation history.  Their solution (from their Figure~12), 
converted to our adopted cosmology, is shown as the green solid curve in
Figure~\ref{fig:OmegaStar_a}, while the shaded region shows their 
95\% confidence interval.   The result and its uncertainty range
are in impressively good agreement with the data (and their uncertainties)
at all redshifts, providing encouragement that these different observables
can be united within a common framework for galaxy mass assembly 
at $0 < z < 3$.

\begin{figure}
\epsscale{0.9}
\plotone{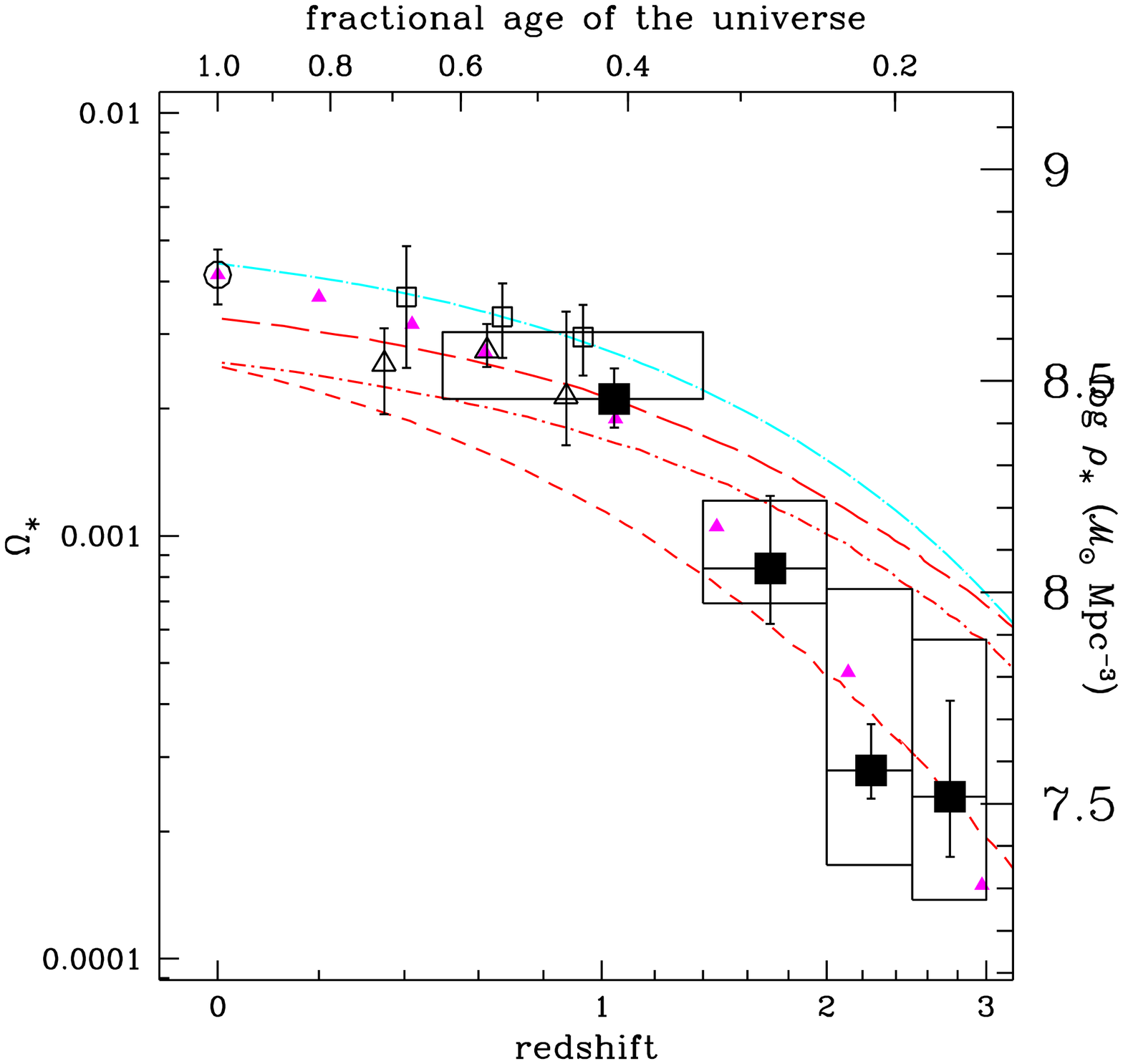}
\caption{
Same as Figure~\ref{fig:OmegaStar_a}, with curves showing 
$\Omega_\ast(z)$ from semi--analytic galaxy evolution 
models by \citet{col00} 
(cyan dot--long-dashed) and \citet{som01} 
(red long--dashed:  ``collisional starburst'';  
dot--short-dashed:  ``accelerated quiescent'';  
short--dashed: ``constant quiescent'').
The small, filled magenta triangles show results from combined
$N$--body and semi--analytic models by \citet{kau99}, rescaled
to match the observed $z = 0$ mass density.  As discussed in
the text, these are probably lower limits to the total
$\Omega_\ast$ at high redshift.
}
\label{fig:OmegaStar_b}
\end{figure}

Figure~\ref{fig:OmegaStar_b} reproduces the data points from Figure~\ref{fig:OmegaStar_a},
superimposing results for $\Omega_\ast(z)$ from recent semi--analytic models 
of galaxy evolution by \citet{col00} and \citet{som01}.  These models use the 
same cosmological parameters as we have adopted here.  The calculations 
of \citet{col00} used the IMF of \citet{ken83}.  We have approximately
rescaled to a Salpeter IMF by multiplying $\rho_\ast(z)$ by a factor of 2.
We also shows results from the $N$--body plus semi--analytic models of 
\citet{kau99}.  Because of their mass resolution, these model outputs 
(taken from the Virgo Consortium web site) are limited to stellar 
masses $\mathM > 10^{10} \mathcal{M}_\odot$ at $z > 1.4$ and 
$\mathM > 2 \times 10^{10} \mathcal{M}_\odot$ at $z < 1.4$.  This is 
similar to the characteristic stellar mass of LBGs at high redshift, 
and thus a large extrapolation would be required to estimate the total 
$\rho_\ast$.  We fit the slope of the differential mass function in the 
models and integrated to compute the stellar mass density 
$\rho_\ast(>10^{10} \mathcal{M}_\odot)$ at all redshifts, but have not 
extrapolated further.   \citet{kau99} assume the IMF of \citet{sca86}, 
requiring an additional adjustment, possibly redshift dependent, for comparison 
to our assumed Salpeter IMF.  Very roughly, we have accounted for this, as 
well as for the extrapolation to ``total'' integrated $\Omega_\ast$, by simply 
rescaling the model $\rho_\ast$ so that it matches the value from the 2dFGRS 
at $z=0$.  

In the model of Cole \etal\ and the favored ``collisional starburst'' 
model of Somerville \etal, the peak redshift for star formation is at 
$z \approx 2.5$ to 4.  Hence, these models show somewhat flatter evolution 
of $\Omega_\ast(z)$, with a larger fraction of the stars in the universe 
formed at $z > 1.5$ than our data suggest.  \citet{kul02} have noted that 
some semi--analytic models also overpredict the mean metallicity in damped 
Lyman~$\alpha$ absorption line systems at high redshift, for similar reasons.
Star formation in the ``constant quiescent'' model of Somerville \etal\ 
peaks at lower redshift ($z \approx 1.5$), yielding better agreement 
with the high redshift data, but underpredicting $\Omega_\ast$ 
at $0 < z < 1.5$.    The model of \citet{kau99} (after rescaling to match
the data at $z=0$) appear to match the shape of $\Omega_\ast(z)$ somewhat better 
than do the other models.  However, given the shapes of the mass functions in 
their simulations, one would expect that a proper extrapolation to total 
$\rho_\ast$ would have a larger effect at higher redshifts.  This would most 
likely flatten the $\Omega_\ast(z)$ evolution somewhat, and therefore the 
higher redshift points from the Kauffmann et al.\ models should probably be 
regarded as lower limits.

Broadly speaking, our estimates of $\Omega_\ast(z)$ favor a global star 
formation history whose rate peaks at $1 < z < 2$ and is smaller at 
higher redshift.   A higher peak redshift for SFR($z$), or a flat 
rate out to $z \gg 3$, would appear to be inconsistent with our measurements,
producing too many stars at $2 < z < 3$ relative to the mass density at lower 
redshifts.  \citet{fer02} have noted a conflict between the stellar mass 
content inferred for $z \sim 3$ LBGs and the evidence for substantial amounts 
of star formation at higher redshifts, as seen directly in the $z \sim 4$ 
LBG population, or as implied by the radiation required to reionize the 
Universe at $z > 6$.  A constant rate of star formation at $3 < z < 6+$ 
would overpredict the stellar masses of LBGs at $z \approx 3$, as well 
as $\Omega_\ast$ at that redshift.  

Neither the current data, nor the theoretical predictions, are yet
sound enough to imply a definite paradox.  If necessary, however, there 
may be several ways to reconcile our measurements of $\Omega_\ast(z)$ 
with evidence for larger star formation rates at higher redshift.  First, 
we might have underestimated luminosity or mass densities at high redshift.
For example, \citet{lan02} have suggested that failure to account for
surface brightness dimming has led to underestimates in the UV luminosity density,
and hence the star formation rate, at high redshift.  Although we stress that 
we have not used isophotal photometry here, virtually any photometry scheme
is subject to some biases, and we might nevertheless have underestimated 
the near--IR fluxes (and hence masses) of high--$z$ objects.  We may also
be missing light from a population fainter than our NICMOS detection limits,
either due to a steep upturn in the luminosity function, or because of 
a steep increase in $\mathM/L$ at fainter luminosities.  We note that surface 
brightness biases would quite probably affect estimates of both SFR($z$) and 
$\rho_\ast(z)$ in similar ways, so that the inconsistency might not be 
significantly narrowed even if corrections were made for missing light.   
The results on $\rho_\ast(z)$ do not allow much room for substantially 
higher rates of star formation at high redshift, either due to surface 
brightness effects or dust obscuration, unless the stellar mass estimates 
are also grossly underestimated.

It is also possible that our assumption of a constant IMF 
at all redshifts could be wrong.  One solution proposed by 
Ferguson \etal\ is to tilt the IMF toward massive star 
formation at higher redshifts.  With a flatter IMF, the massive stars 
that would produce UV radiation at the reionization epoch, or from 
LBGs at $z \approx 4$, are accompanied by a smaller total 
formation rate of stellar mass, which is dominated by lower--mass, 
long--lived stars.  In this way, we may reconcile the UV output at
$z \gg 3$ with the relatively low $\Omega_\ast$ seen at $2 < z < 3$.
Eventually, however, there must be a change in the IMF to produce
the longer--lived stars which dominate the stellar mass build--up 
at $1 < z < 3$.   It is not clear how to realistically incorporate 
a time-- or redshift--dependent IMF into current models of galaxy
evolution, but this may become necessary.

It is also interesting to note the relative paucity of galaxies with 
$\mathM > 10^{11} \mathM_\odot$ at almost any redshift in the HDF--N
(see Figure~\ref{fig:Mass1vsVz}).  \citet{col01} find a characteristic stellar 
mass of present--day galaxies\footnote{
In passing, we note that \citet{col01} adopt SFR models which start at 
$z = 20$ when fitting the photometry for 2dFGRS galaxies and calculating 
stellar masses.  It is therefore possible that their masses are overestimated 
if, in fact, the bulk of stars formed at lower redshifts, as suggested by our 
observations.  However, we also note that their characteristic mass estimate 
$\mathM^\ast$ agrees reasonably well with that from \citet{kau02} (after
rescaling for differences in the assumed IMF), who allowed greater variety 
when modeling galaxy ages and star formation histories.
}
$\mathM^\ast = 1.4 \times 10^{11} \mathM_\odot$.
There are only a handful of HDF--N galaxies at $0.5 < z < 2$ which exceed 
this mass, despite adequate total volume to have found roughly a dozen
in the local universe.  This is true even when the maximum $\mathM/L$
2--component star formation models are adopted for the HDF--N galaxies.
This might be attributable to small number statistics and to the small
co--moving volume of the HDF--N.  The most massive galaxies are rare,
and moreover they may be expected to cluster strongly, making it still more
difficult to take a fair census from a single sight line, small--volume
survey.  However, it might also indicate that the most massive 
galaxies continued to build up their stellar content at relatively late 
epochs, either by merging or late star formation, even after the time when 
much of the overall density of stars in the universe had already formed.

Although we cannot yet derive robust stellar mass functions at 
high redshift from our data, the qualitative behavior seen in 
Figure~\ref{fig:Mass1vsVz} is reminiscent of that found in recent 
cold dark matter models of galaxy formation in a $\Lambda$--dominated 
universe.  \citet{bau02} show stellar mass functions at various redshifts
derived from the models of \citet{col00}, and similar results can be
extracted from the models of \citet{kau99} available from the Virgo 
Consortium web site.   Locally, galaxies with stellar masses 
$\mathM \approx 10^{11} h_{70}^{-2} \mathcal{M}_\odot$ have a space 
density $\phi \approx 2\times 10^{-3} h_{70}^{-3}$~Mpc$^{-3}$.
We consider objects at a fixed space density corresponding to 
that of $10^{11} \mathM_\odot$ galaxies at $z = 0$.  In both sets 
of models, at $z \approx 1$, 2, and 3, the stellar mass of objects 
with this characteristic space density has decreased by factors of 
$\sim 2$, 4, and 10, respectively.  At $z = 3$, the characteristic
mass is approximately $10^{10} \mathM_\odot$, as found by PDF.
The models differ in their predictions for the space density of 
$10^{11} \mathM_\odot$ objects, with \citet{bau02} finding more
rapid evolution.  This is unsurprising in the exponential part
of the mass function, where small differences along the mass axis
correspond to very large differences in space density.  Both models, 
however, predict that $\phi(10^{11} \mathM_\odot)$ changes by more 
than an order of magnitude between $z = 2$ and 1, and that galaxies 
with that mass largely disappear by $z = 3$, with a space density 
less than 1\% of that locally.

Although we find no HDF--N galaxies at $z > 2$ with 
$\mathM > 10^{11} \mathcal{M}_\odot$, this does not mean that
such galaxies do not exist in the Universe.  \citet{sha01} find 
a small number of LBGs with inferred 
$\mathM \gtrsim 10^{11} \mathM_\odot$ among a sample taken 
from their large, ground--based survey.   Moreover, there are 
powerful radio galaxies in this redshift range with substantially 
brighter $K$--band magnitudes, and presumably larger stellar 
masses, than anything in the HDF--N (c.f.\ \citet{deb02}).
Overall, however, galaxies with these large stellar masses at
$z \approx 3$ appear to be quite rare.  Wide--field (but deep) 
near--infrared surveys with reliable redshift information will 
be needed to quantify this properly.

From large redshift surveys of $K$--selected galaxy samples, 
\citet{coh02} and \citet{cim02b} have argued that the galaxies 
which make up the bulk of the stellar mass distribution today have 
evolved in a fashion that is broadly consistent with pure 
luminosity evolution, at least from $z \approx 1$ or 1.5 to the 
present.  We see little evidence to contradict this from the 
HDF--N.  The difference in luminosity density from $z \approx 1$ to 
the present is broadly consistent with ``passive'' luminosity
evolution ($\sim 0.85$~mag in rest--frame $L^\ast_B$) with only 
modest ($\sim 25\%$) density evolution (see \S\ref{section:results}).
At higher redshift, however, our results would strongly disagree
with simple prescriptions for pure luminosity evolution.

\subsection{Field--to--field variations
\label{section:field2field}}

Our present analysis is limited to one very deep, but very small, 
field of view, and is therefore subject to uncertainties due to the
relatively small co--moving volume and the effects of galaxy clustering.
We have quantified the shot--noise uncertainties, but we have made 
no attempt to deal with the possible influence of clustering, 
other than to use very wide redshift intervals when analyzing 
the data in the hope of averaging over line--of--sight variations.
Although the co--moving volume of each of our redshift bins is 
roughly equal, the co--moving, line--of--sight path--length ranges 
from $2300 h_{70}^{-1}$~Mpc at $0.5 < z < 1.4$ to $530 h_{70}^{-1}$~Mpc 
at $2.5 < z < 3$.  We know that the redshift distributions of 
galaxies at both low and high redshift exhibit ``spikes,'' 
as pencil--beam surveys penetrate groups and sheets along 
the sight--line (cf.\ \citet{lef96}, \citet{coh96}, \citet{ade98}).  
We would expect our redshift bins to average over several of these 
overdensities and the ``voids'' between them.  Nevertheless, we may 
expect that clustering will introduce some additional variance, 
beyond that included in our error estimates.  Indeed, it may
be expected that the most massive galaxies cluster most strongly
at any redshift.  

One motivation for the Hubble Deep Field South project
(HDF--S, \citet{wil00}) was to provide a cross--check on results
from the HDF--North, as insurance against ``cosmic variance''.
\citet{rud01} have presented deep, near--infrared observations
of the HDF--S from the VLT.  To date, few spectroscopic redshifts
are available in the HDF--S, but Rudnick \etal\ have estimated 
photometric redshifts and rest--frame optical luminosities for 
galaxies in their sample.   They find 6 galaxies in the HDF--S 
with $2 < z_{phot} < 3$ and $L_B > 10^{11} h_{70}^{-2} L_\odot$ 
whereas we find two in the HDF--N.  These differences may be within 
the range expected from statistical fluctuations, which have 
already been taken into account by our bootstrap error analysis 
of the luminosity density, although larger variance due to clustering
(particularly for the most luminous objects) are to be expected.
The most luminous HDF--S galaxies at $2 < z_{phot} < 3$ 
are about twice as bright as those in the HDF--N.  
It is important to remember, however, that for distributions 
like the Schechter function, the objects that contribute most 
to the total luminosity or mass density
are those around $L^\ast$ or $\mathM^\ast$, i.e., the 
``typical'' objects, not the brightest ones.  
\citet{lab02} identify 7 HDF--S galaxies with 
$1.95 < z_{phot} < 3.5$ and with $V-H$ colors redder
than galaxies at similar redshift in the HDF--N 
(cf.~PDF Figure 1).  If the HDF--S photometric redshift estimates
are correct, then these galaxies may have higher $\mathcal{M}/L$ 
than the large majority of galaxies with similar redshifts 
and $B$--band luminosities in either the northern or southern fields.

The differences between the HDF--S and HDF--N may be actually be 
greater at $0.5 < z < 2$, where \citet{rud01} find no galaxies 
with $L_B > 10^{11} h_{70}^{-2} L_\odot$, 
compared to six in the HDF--N.  The fact that the HDF--N luminosity 
density at $z \approx 1$ agrees well with that measured in other 
surveys that cover wider fields or multiple sight lines
gives us some confidence in our mass density estimates 
at that redshift.  At $1.4 < z < 2$, where both samples rely 
very heavily on photometric redshifts, Rudnick \etal\ find only
two HDF--S galaxies with $L_B > 2 \times 10^{10} h_{70}^{-2} L_\odot$ 
(roughly present--day $L_B^\ast$), compared to 17 in the 
HDF--N.  Overall, these comparisons point to the need for 
caution interpreting results from any one deep, narrow
survey, and highlight the importance of wider fields
of view and multiple sight lines.

\subsection{Photometric redshift errors
\label{section:photzerrs}}

Many of the galaxies in the HDF--N have only photometric redshift 
estimates.  It is therefore important to consider the effects 
of photometric redshift errors on our results.
At all redshifts, the brighter galaxies make up a substantial 
fraction of the total stellar mass density, and in the HDF--N 
most of these have spectroscopic redshifts.  For the range of 
model metallicities and star formation histories that we have 
considered, the fraction of $\log \rho_\ast$ 
that is made up by objects with spectroscopic redshifts in 
each of our redshift bins is:  
$0.5 < z < 1.4$ --  74-78\%; $1.4 < z < 2.0$ --  35-43\%; 
$2.0 < z < 2.5$ --  22-28\%; $2.5 < z < 3.0$ --  32-41\%.
Photometric redshift errors are therefore unlikely to have 
a dominant effect on mass densities at $z \approx 1$, but may 
be more important at higher redshifts.

Photometric redshift errors come in two ``flavors'':  small uncertainties,
and catastrophic outliers (\S\ref{section:data}).  To first order, 
small redshift errors translate quadratically to changes in 
luminosity and hence to the derived masses.  We have perturbed 
our photometric redshifts by amounts drawn from the observed 
$\Delta_z$ distribution for the spectroscopic sample, truncated 
at $|\Delta_z| < 0.1$.  Objects with spectroscopic redshifts 
were not perturbed.  The changes in $\log \rho_L$ within 
each redshift bin are small.  Averaging over many perturbations, 
the mean and RMS offsets range from $-0.006 \pm 0.003$~dex at 
$0.5 < z < 1.4$ to $+0.05 \pm 0.03$~dex at $2.5 < z < 3$, and 
are smaller than other computed uncertainties.

A greater concern is the possibility of catastrophic 
outliers.  The most common cause is aliasing between minima
in $\chi^2(z_{phot})$, e.g., from confusion between different 
breaks or inflections in the spectral energy distribution.
In such cases, it is possible to mistake a faint, low--redshift 
object for a very luminous one at high redshift.  Moreover, 
the rest--frame colors, and hence $\mathM/L$ derived
from spectral model fitting, could be completely wrong.
A moderately red galaxy at low redshift, if incorrectly assigned 
a large $z_{phot}$, might be fit with a very red, high--$\mathM/L$ 
stellar population, or we might miss a massive, red, high--$z$
galaxy if it were erroneously placed at low $z_{phot}$.  
The impact of such errors is largest at higher redshifts.
Transferring a few galaxies between (say) $z_{phot} < 1.4$ and 
$2 < z < 3$ will have relatively little effect on global quantities
at low redshift (where their inferred luminosities would be small), 
but might cause a significant change in the high redshift bins.  

It is difficult to assess the impact of such catastrophic redshift 
errors formally, but we have looked closely at the situation in 
the HDF-N, and offer the following observations.  The largest
effect would most likely be due to redshift errors for galaxies
with relatively bright near--infrared magnitudes.  Only 18 (out 
of 107) HDF--N galaxies with $\nich < 23$ lack spectroscopic redshifts.
These 18 galaxies all have best--fitting $0.9 < z_{phot} < 1.8$.
Examining the photometric redshift likelihood function for each,
we find that four have next--best values in the range $2 < z_{phot} < 3$.
In three of the four cases, $\chi^2$ for the next--best $z_{phot}$ 
is 45 to 125$\times$ poorer than that of the best fit.  The morphologies 
and SED shapes strongly suggest that these are ordinary, early--type 
galaxies at $1.1 < z < 1.2$.  One galaxy (4-186)
is more problematic.  It is best fit by $z_{phot}=1.71$
(\citet{fer99} find a very similar $z_{phot} = 1.60$), but
$\chi^2$ is only $2.5\times$ worse at $z_{phot}=2.86$.   
We see no reason to doubt the best--fit value, but this is an 
example of a galaxy which might, in principle, be at higher redshift.
With $\nich = 22.2$ and red colors, its stellar mass would be much 
larger if placed at the higher redshift.  At $23 < \nich < 24$ there 
are five more galaxies with best--fit $z_{phot} < 2$ and next--best 
$2 < z_{phot} < 3$.  For three, the next--best $\chi^2(z_{phot})$ 
values are again far worse, and careful inspection makes $z > 2$ 
seem unlikely.  Two others have smaller $\chi^2$ ratios, and
for one (3-515.11), both possibilities 
($z_{phot} = 1.24$ or 2.64) seem comparably likely, with
$\chi^2$ differing by a factor of only 1.2.  At the larger 
redshift, this would be a fairly ordinary, $\sim L^\ast$ 
LBG with blue colors and typical $\mathM/L$, and would only
slightly increase the global $\rho_M$.   At still fainter magnitudes,
a modest rate of catastrophic errors would have very little effect 
on the global mass densities.

In another test, we compared two somewhat different photometric 
redshift fits:  one (the default for this paper) using the 
WFPC2+NICMOS+IRIM photometry and ``trained'' spectral templates 
from \citet{bud00}, versus another which excludes the ground--based 
$K$--band data and uses more conventional templates based on the 
\citet{col80} empirical SEDs.  Only one blue, faint galaxy shifted 
into or out of the range $2 < z_{phot} < 3$, with negligible impact 
on luminosity or mass densities.  Finally, our photometric redshift 
estimates can optionally incorporate a constraint on rest--frame 
$B$--band luminosity, somewhat similar to the Bayesian priors 
used in the method of \citet{ben00}.
This might, in principle, cause erroneous $z_{phot}$ estimates 
for unusually luminous galaxies at high redshift.  However, 
when we switch off this luminosity prior, no galaxy is 
reassigned to $2 < z_{phot} < 3$, and there is no significant 
effect on our results.   

Without reliable spectroscopic redshifts for all galaxies, we 
cannot be absolutely certain that photometric redshift errors 
do not affect our derived luminosity or mass densities.  
However, from the tests performed here we see no strong evidence
that this is a significant problem for the HDF--N sample.  
It is important, however, to remember that for some questions, 
a comparatively small error rate in photometric (or spectroscopic) 
redshifts can introduce potentially large errors in the derived 
results.  Other problems (e.g., galaxy clustering) are more 
robust against modest redshift error rates.

\section{Conclusions
\label{section:conclusions}}

We have used deep near--infrared imaging of the Hubble Deep Field North
to select a sample of galaxies based on their optical rest--frame light
out to $z = 3$, and have estimated their stellar mass content following
the procedures outlined by \citet{pap01}.  We find that the rest--frame 
$B$--band co--moving luminosity density is somewhat larger at high redshift 
than the present--day value, although constant to within a factor of 3 over 
the redshift range $0 < z < 3$.  However, galaxies at higher redshifts
are much bluer than most luminous and massive galaxies locally, and 
the mean mass--to--light ratio evolves steeply with redshift.  The result 
is that the global stellar mass density, $\Omega_\ast$, rises rapidly
with cosmic time, going from 3--14\% of the present--day value at 
$z \approx 2.7$, reaching 50--75\% of its present value by $z \approx 1$, 
and then (based on results from the literature) changing little to the 
present day.  In broad terms, the concordance between previous estimates 
of the cosmic star formation history and these new measurements of its 
time integral strongly suggests $1 < z < 2.5$ as a critical era, when
galaxies were growing rapidly toward their final stellar masses.
There are hints of a possible inconsistency between the global 
mass density $\rho_\ast$ at $z \approx 3$ and evidence for 
high star formation rates at $3 < z < 6$.  One solution to this 
might be to tilt the IMF toward high--mass star formation at
early times, as suggested by \citet{fer02}.

\citet{ren99} has recapped the evidence and arguments which suggest
that spheroids contain $\gtrsim 30$\% of the stars in the local universe,
and that those stellar populations are old and formed at high redshift. 
Recent estimates from the SDSS place $\gtrsim 50\%$ of the local stellar 
mass density in galaxies with red colors and/or large 4000\AA\ break 
amplitudes \citep{hog02,kau02}.  Other estimates have set even larger 
values for the stellar mass fraction in galaxy spheroids (e.g., $\sim 75\%$ 
from \citet{fuk98}).  This would appear to run counter to the evidence 
presented here, which suggests that $< 20$\% of the present--day 
stellar mass was in place at $2 < z < 3$.  

Our estimates of $\Omega_\ast(z)$ are not necessarily inconsistent with 
the ages of the red stellar populations which contribute so substantially 
to $\Omega_\ast$ today.  The presently favored world model dominated by 
a cosmological constant allows more time to elapse at lower redshifts 
for a given value of $H_0$.  In this case, the old ages of stars in 
local spheroids are more comfortably accommodated with lower formation 
redshifts, $z \approx 2$, where the look--back time is 
$\approx 10 h_{70}^{-1}$~Gyr.  We have found that most (at least half) 
of the present--day stellar mass was in place at $z \approx 1$ (corresponding 
to a lookback time of $7.7h_{70}^{-1}$~Gyr), when the cosmic ``building boom'' 
seems to have been winding down.  It may be that the local amount of old, 
red starlight can be comfortably accommodated in such a universe.  

A census of stellar populations in galaxies at $z \approx 1$ may 
provide a stronger test of our HDF--N results at $z > 2$.  
\citet{cim02a} have analyzed composite spectra of red 
galaxies at $z \approx 1$, and have argued that a significant fraction 
of these galaxies are dominated by stellar populations with a minimum
age of $\sim 3$~Gyr.  Such stars therefore would have formed at
$z \gtrsim 2.3$ for the cosmology adopted here.  There is not
yet a good estimate of the total, co--moving mass density in old 
stars at $z \approx 1$, but if it were substantial this might
prove to be discrepant with evidence for a rapid mass 
build--up from $z \approx 3$ to 1.  If necessary, it might be possible 
to push our estimate of $\Omega_\ast$($z > 2$) upward toward 30\% of
the present--day value if each galaxy were assigned its maximum stellar 
mass (at, say, 68\% confidence) allowed by our 2--component star 
formation models -- e.g., if indeed there were a very early epoch 
of rapid star formation preceding the redshift range studied here.
However, once again we note that these models seem {\it ad hoc} 
and somewhat unphysical.

The robustness of these results is limited by several factors.
First, there is the small volume of the HDF--N, and the vulnerability
of a single sight--line to the effects of large scale structure,
which we have discussed in \S\ref{section:field2field}.  Our results 
at $z \sim 1$ generally agree well with other estimates of the 
luminosity and stellar mass density that can be found in the 
literature.  At higher redshifts, there is some evidence for 
differences between the HDF--N and HDF--S sight lines.  It is not
yet clear if the implied stellar mass densities at high redshift
in these two fields disagree by more than our estimated uncertainties
for the HDF--N alone.  Second, photometric data reaching to the 
$K$--band can sample only rest--frame optical wavelengths at 
$z \sim 3$, where the estimates of $\mathM/L$ depend strongly on 
assumptions about the stellar population metallicity and past star 
formation history.  Also, at the highest redshifts, even the NICMOS 
data do not reach far enough down the galaxy 
luminosity function to constrain the total luminosity or stellar mass 
densities as tightly as one would like, leaving significant uncertainty 
on these integrated quantities.  However, if we are missing most of
the total stellar mass density at $2 < z < 3$ for this reason, then 
it must be found primarily in low--luminosity galaxies.  This would 
strongly contrast with the situation today and at intermediate redshifts, 
where the bright end of the luminosity function dominates the
integrated stellar mass density.  Finally, we note that our choice
of cosmology does affect the redshift dependence of $\Omega_\ast(z)$.
In a flat, matter dominated universe with the same value of $H_0$, 
the luminosity densities would be 80\% larger at $2.5 < z < 3.0$.
However, the {\it ratio} of densities between $z \approx 1$ and 
$z \approx 2.7$ would be change by only 15\%, since the change in
world model would affect all of the high redshift points.  Therefore,
the general conclusion of a steep increase in $\Omega_\ast$ at $1 < z < 3$
would be essentially the same.

The launch of the {\it Space Infrared Telescope Facility (SIRTF)} 
should help to tie down some of these loose ends.  In particular, 
the Great Observatories Origins Deep Survey (GOODS, \citet{dic02}) 
incorporates a {\it SIRTF} Legacy program which will collect the deepest 
observations with that facility at 3.6--24$\mu$m, in part to address 
this very issue.  The GOODS program should help to improve the situation 
in several ways.  It will survey two independent sight--lines, covering 
a solid angle $60\times$ larger than that considered here.  {\it SIRTF} 
observations out to $8\mu$m will sample rest--frame $2\mu$m light from 
galaxies out to $z = 3$, and $1\mu$m light to $z = 7$.  This should 
should provide much better constraints on stellar masses than do our 
current NICMOS and ground--based data, which only reach optical 
rest--frame light at high redshift.  {\it SIRTF} is a comparatively 
small telescope (85~cm aperture), and its sensitivity to the faintest 
galaxies will be limited by collecting area and source confusion.  
GOODS is designed to robustly detect typical ($\sim 10^{10} M_\odot$) 
Lyman break galaxies at $z = 3$ (and to push to fainter luminosities 
and higher redshifts over a smaller solid angle in its ultradeep survey).
But just as we found with our NICMOS and ground--based data here, 
estimates of the total, integrated mass density may be limited by 
poor constraints on the faint--end slope of the galaxy luminosity 
function at these large redshifts.  Ultimately, observations from 
the {\it James Webb Space Telescope}, which will reach nJy sensitivities 
with observations out to at least $5\mu$m, may be required to fully 
address this problem, and to push toward still higher redshifts 
and earlier epochs.

\acknowledgements

We would like to thank the other members of our HDF--N/NICMOS GO team
who have contributed to many aspects of this program, and the STScI
staff who helped to ensure that the observations were carried out in
an optimal manner.  We thank Matthew Bershady, Peter Eisenhardt, 
Richard Elston, and Adam Stanford, who helped collect the ground--based 
$K$--band data,  Stephane Charlot and Gustavo Bruzual for 
providing an up--to--date version of their population synthesis code, 
Kurt Adelberger and Chuck Steidel for permission to use several 
unpublished HDF--N redshifts, Christian Wolf for tabulated data
from the COMBO-17 program, and Adam Riess for statistical advice.
Alvio Renzini and Mike Fall provided valuable comments on early
drafts of this manuscript, and we also benefited from discussions
with Jim Peebles and Andrea Cimatti.  The anonymous referee 
contributed several very helpful suggestions which have improved
this paper.   Support for this work was provided by NASA through 
grant number GO-07817.01-96A from the Space Telescope Science Institute, 
which is operated by the Association of Universities for Research 
in Astronomy, Inc., under NASA contract NAS5-26555.

\clearpage

\begin{deluxetable}{lcllcll}
\tablewidth{0pt}
\tablecaption{HDF-N rest--frame $B$--band luminosity densities
 ($\Omega_M = 0.3$, $\Omega_\Lambda = 0.7$, $h = 0.7$)\label{tab:lumdenstab}}
\tablehead{
  & \multicolumn{3}{c}{$1/V_{max}$ summation}
  & \multicolumn{3}{c}{Schechter integration} \\

  \colhead{$z$ range}
& \colhead{$\log\rho_L$}
& \colhead{$\pm 68\%$} 
& \colhead{$\pm 95\%$} 
& \colhead{$\log\rho_L$}
& \colhead{$\pm 68\%$} 
& \colhead{$\pm 95\%$} \\

& ($h_{70} L_\odot$~Mpc$^{-3}$) & & & ($h_{70} L_\odot$~Mpc$^{-3}$) & & 
}

\startdata

 0.5--1.4 & $8.49$ & $_{-0.05}^{+0.05}$ & $_{-0.10}^{+0.09}$ & $8.48$ & $_{-0.06}^{+0.06}$ & $_{-0.13}^{+0.12}$ \\
 1.4--2.0 & $8.18$ & $_{-0.06}^{+0.06}$ & $_{-0.13}^{+0.12}$ & $8.25$ & $_{-0.11}^{+0.15}$ & $_{-0.26}^{+0.52}$ \\
 2.0--2.5 & $8.23$ & $_{-0.05}^{+0.04}$ & $_{-0.09}^{+0.08}$ & $8.26$ & $_{-0.05}^{+0.06}$ & $_{-0.11}^{+0.13}$ \\
 2.5--3.0 & $8.11$ & $_{-0.09}^{+0.07}$ & $_{-0.18}^{+0.15}$ & $8.42$ & $_{-0.18}^{+0.70}$ & $_{-0.34}^{+\infty}$ \\
 2.5--3.0\tablenotemark{a} & & & & $8.22$ & $_{-0.12}^{+0.14}$ & $_{-0.24}^{+0.44}$

\enddata

\tablenotetext{a}{Schechter function refit fixing $\alpha = -1.4$.}

\end{deluxetable}

\begin{deluxetable}{lrcccc}
\tablewidth{0pt}
\tablecaption{Mean $B$--band mass--to--light ratios from stellar population fitting\label{tab:moltab}}
\tablehead{
  \colhead{$z$ range} 
& & \multicolumn{4}{c}{$\langle \mathM/L_B \rangle$ (solar units)}\\

& & \multicolumn{2}{c}{$Z = 1.0 Z_\odot$}
& \multicolumn{2}{c}{$Z = 0.2 Z_\odot$} \\

& \colhead{$\Psi(t)$:}
& \colhead{1--comp.}
& \colhead{2--comp.}
& \colhead{1--comp.}
& \colhead{2--comp.} 
}

\startdata

 0.5--1.4 && $0.96_{-0.06}^{+0.10}$ & $1.36_{-0.08}^{+0.13}$ & $1.10_{-0.13}^{+0.11}$ & $1.38_{-0.11}^{+0.12}$ \\ 
 1.4--2.0 && $0.64_{-0.10}^{+0.14}$ & $0.93_{-0.08}^{+0.12}$ & $0.54_{-0.14}^{+0.11}$ & $0.76_{-0.12}^{+0.12}$ \\ 
 2.0--2.5 && $0.21_{-0.02}^{+0.05}$ & $0.57_{-0.07}^{+0.11}$ & $0.13_{-0.02}^{+0.03}$ & $0.35_{-0.05}^{+0.06}$ \\
 2.5--3.0 && $0.20_{-0.03}^{+0.11}$ & $0.48_{-0.08}^{+0.18}$ & $0.12_{-0.03}^{+0.07}$ & $0.27_{-0.05}^{+0.14}$ 

\enddata

\end{deluxetable}

\begin{deluxetable}{lrcccc}
\tablewidth{0pt}
\tablecaption{Stellar mass densities
 ($\Omega_M = 0.3$, $\Omega_\Lambda = 0.7$, $h = 0.7$)\label{tab:rhoMtab}}
\tablehead{
  \colhead{$z$ range} 
& & \multicolumn{4}{c}{$\log \rho_\ast$ ($h_{70} \mathM_\odot$~Mpc$^{-3}$)}\\

& & \multicolumn{2}{c}{$Z = 1.0 Z_\odot$}
& \multicolumn{2}{c}{$Z = 0.2 Z_\odot$} \\

& \colhead{$\Psi(t)$:}
& \colhead{1--comp.}
& \colhead{2--comp.}
& \colhead{1--comp.}
& \colhead{2--comp.} 
}

\startdata

0.5--1.4 && $8.46_{-0.07}^{+0.07}$ & $8.61_{-0.07}^{+0.07}$ & $8.52_{-0.08}^{+0.07}$ & $8.62_{-0.07}^{+0.07}$ \\ 
1.4--2.0 && $8.06_{-0.13}^{+0.17}$ & $8.22_{-0.12}^{+0.16}$ & $7.97_{-0.17}^{+0.17}$ & $8.12_{-0.13}^{+0.16}$ \\ 
2.0--2.5 && $7.58_{-0.07}^{+0.11}$ & $8.01_{-0.08}^{+0.09}$ & $7.36_{-0.08}^{+0.11}$ & $7.79_{-0.09}^{+0.10}$ \\ 
2.5--3.0 && $7.52_{-0.14}^{+0.23}$ & $7.89_{-0.15}^{+0.20}$ & $7.27_{-0.18}^{+0.27}$ & $7.65_{-0.15}^{+0.22}$ 

\enddata

\end{deluxetable}

\end{document}